\documentclass[usenatbib,usegraphicx,useAMS,fleqn]{mn2e}

\usepackage{amsmath,amssymb}

\title{Resonant Excitation of White Dwarf Oscillations in Compact Object Binaries:
       1. The No Back Reaction Approximation}
\author[Yasser Rathore, Roger D. Blandford and Avery
  E. Broderick]{Yasser Rathore,$^{1}$\thanks{yasser@caltech.edu}
	Roger D. Blandford$^{1,2}$ and Avery E. Broderick$^1$ \\ \\
	$^1$Theoretical Astrophysics, Caltech 130-33, Pasadena, CA 91125, USA \\
	$^2$Kavli Institute for Particle Astrophysics and Cosmology, Stanford University, Stanford, CA 94305, USA}
\date{Accepted 2004; Received 2004}
\pagerange{\pageref{firstpage}--\pageref{lastpage}}
\pubyear{2004}

\begin{document}
\maketitle
\label{firstpage}
\begin{abstract}
We consider the evolution of white dwarfs with compact object companions 
(specifically black holes with masses up to 
$\sim10^6$~M$_\odot$, neutron stars, and other 
white dwarfs). We suppose that the orbits are initially quite elliptical
and then shrink and circularise under the action of gravitational radiation.
During this evolution, the white dwarfs will pass through resonances
when harmonics of the orbital frequency match 
the stellar oscillation eigenfrequencies.
As a star passes through these resonances, the associated
modes will be excited and can be driven to amplitudes that are so large that 
there is a back reaction on the orbit which, in turn, limits the growth of the
modes. A formalism is presented for describing this dynamical interaction
for a non-rotating star in the linear approximation when the orbit can be
treated as non-relativistic.
A semi-analytical expression is found for computing the resonant energy
transfer as a function of stellar and orbital parameters for the regime
where back reaction may be neglected. This is used to calculate the results of
passage through a sequence of resonances for several hypothetical systems.
It is found that the amplitude of the $\ell=m=2$ \emph{f}-mode can be driven
into the non-linear regime for appropriate initial conditions.
We also discuss where the no back reaction approximation is expected to fail,
and the qualitative effects of back reaction.
\end{abstract}
\section{Introduction}
White dwarfs, the normal evolutionary 
endpoint for stars less massive than $\sim8$~M$_\odot$, are extremely
common; the halo of our Galaxy contains several billion of them.
They are observed frequently in binary systems, with normal 
stellar companions, as cataclysmic variable stars
and, less often, with compact object companions, WDCO systems.
Many of these systems are produced naturally in binary star evolution
and, as a consequence, have circular orbits. However, it is 
also possible to form eccentric, WDCO binaries following
stellar capture or exchange in a dense stellar environment.
In the present paper, we do not consider the formation of such
binaries in detail; we list some possible formation mechanisms, but
our detailed considerations are restricted to the evolution of such
systems after they form.

There are three possible types of WDCO systems. Perhaps the 
most common are double degenerates \citep[e.g.][]{max02}, which have been
observed to have periods as short as $\sim5$~min \citep{ram02}.
A second class is white dwarf-neutron star binaries, of which there
are over 50 known or suspected \citep{lor01}, with orbital periods as short
as 3 hr \citep{edw00}. The third class are white dwarf-black hole binaries.
For black hole companions in the stellar mass range, there are no known
examples--they would be extremely hard to find. There is, however, growing
theoretical evidence for, and observational evidence consistent with,
the existence of intermediate mass black holes (IMBHs).
X-ray observations of nearby galaxies hint at the possibility
of a large population of black holes in the
$\sim10^2$--$10^4$~M$_\odot$ mass range \citep{col99,col02}. The
interpretation of these observations is the source of some debate,
but accreting IMBHs are certainly a plausible explanation.
These IMBHs may have been formed in the early universe from 
$10^2$--$10^3$~M$_\odot$ stars \citep{bro99}. Also, recent dynamical simulations of
dense stellar clusters have demonstrated that runaway collisions
can lead rapidly to the formation of black holes with masses in the
$\sim10^2$--$10^3$~M$_\odot$ range \citep{zwa04}. More generally, the
population of IMBHs could span the entire interval
between stellar and massive black holes  ($\sim10^2$--$10^6$~M$_\odot$).
They could either tidally capture white dwarfs from the field \citep{fab75,pre77},
or participate in three-body processes in stellar clusters and end up in
bound orbits with white dwarf companions \citep{hil77}.

The tidal capture mechanism is simple in principle. 
When the radius of periastron is no more than a few times the
Roche limit, the normal modes of the white dwarf can be excited
non-resonantly. On egress, the energy of the oscillation is taken
out of the orbit, which can become bound. The energy
transfer need be no more than
$\sim$($30$--$100$~km~s$^{-1}$)$^2 \equiv 10^{13-14}$~erg~g$^{-1}\equiv10$--$100$~eV
per nucleon for capture to ensue. Subsequent periastron passages can lead to
additional mode excitation and the possibility of heating the star if the
oscillation energy is thermalised. For a black hole companion, this mechanism
requires that the Roche limit lie not too far inside the event horizon, which
sets an upper limit of $\sim10^6$~M$_\odot$ on the black hole mass. 
(When the black hole mass exceeds $\sim3\times10^5$~M$_\odot$, 
it turns out that gravitational bremsstrahlung is more important than tidal
capture and contributes a larger capture cross-section.)

Whatever their detailed nature and origin, WDCO binaries evolve
dynamically under the action of gravitational radiation. 
The orbital periods and eccentricities 
will decrease until the former reaches the Roche period, 
$\sim10-100$~s depending upon mass, when the star will be torn
apart by tidal forces. During inspiral, a WDCO system will pass through a
series of resonances between harmonics of the orbital frequency and the
white dwarf normal mode eigenfrequencies. Typically, the system will
spend many orbits near each resonance, and consecutive resonances for
a given mode will be separated by a much larger number of orbits
associated with the gravitational inspiral time-scale. Passage through
a sequence of such resonances will result in transfer of
energy from the orbit to the oscillations and may drive the amplitudes
of the oscillations non-linear, with possibly observable consequences.

If the energy transfer from the orbit to the stellar oscillation
is as large as $\sim10$~keV per nucleon and the mode energy is thermalised,
then it is possible to detonate the star, releasing more nuclear energy
($\sim1$~MeV per nucleon) than the gravitational binding energy
($\sim0.1$--$0.3$~MeV per nucleon). Typically, for a black hole
companion, the orbital binding energy will be greater than the nuclear energy
released, so most of the ejecta will be trapped. However, for a white dwarf or
neutron star companion, the ejecta can become unbound.

We note that WDCO binaries also constitute one of the prime potential sources for
LISA \citep[e.g.][]{fin00}. The considerations presented below may be relevant to
gravitational wave observations of these systems.

In order to understand how much energy is transferred from the 
orbit to the white dwarf during inspiral, it is necessary 
to consider the evolution of the system after it 
has lost much of its initial eccentricity. 
We present a general formalism for handling
tidal resonances in WDCO binaries in the linear normal mode approximation.
We restrict ourselves to non-rotating stars because white dwarfs are
observed generally to be slowly rotating, and they are not expected to maintain
corotation during inspiral \citep[c.f.][]{bil92}. However, this needs further
investigation. In addition, we mostly confine our attention to the $\ell=m=2$
\emph{f}-mode because it is expected to be the dominantly excited one. Nonetheless,
our formalism may be applied to other modes as well.
Of particular interest may be $g$-modes, as these have lower
frequencies than the $f$-modes, and can therefore be excited
at fundamental resonance in circular orbits before tidal
disruption.

We begin with a description of the general formalism in \S\ref{S:GF}.
This includes a brief description of the
resonant excitation of a simple harmonic oscillator (\S2.1), a summary
of the structure and normal modes of cold white dwarfs (\S2.2),
an overview of gravitational radiation reaction (\S2.3), and our
formalism for tidal excitation (\S2.4). In \S3, we discuss some
physical aspects of the dynamical problem, including mode damping and time-scales.
We also discuss the physical regimes where various effects are important.
In \S4, we apply the general formalism to calculate the energy transfer for
a resonance in an eccentric orbit, ignoring back reaction. In \S5, we
discuss the regime where this result is expected to be valid, and
describe the qualitative effects of back reaction.
This is followed by a description of the long term evolution
of WDCO binaries in the absence of back reaction.
We present our conclusions in \S6.

\section{General Formalism}\label{S:GF}
\subsection{Simple harmonic oscillator}\label{SS:SHO}
Consider an undamped simple harmonic oscillator
with natural frequency $\omega_0$ and displacement
$x(t)$ subject to an external force per unit mass $F(t)$.
The equation of motion,
\begin{equation}
\label{eq:sho}
\ddot{x} + \omega_0^2 x = F(t) \ ,
\end{equation}
can be easily solved to get
\begin{align}
\dot{x}(t) &= \Re\left[\zeta(t)\right]  \ , \\
x(t) &= \frac{1}{\omega_0}\Im\left[\zeta(t)\right] \ ,
\end{align}
where
\begin{equation} \label{eq:sho_zeta}
\zeta(t) = \zeta_0 e^{i\omega_0 t} + \zeta_1(t) \ ,
\end{equation}
and
\begin{equation} \label{eq:sho_zeta1}
\zeta_1(t) \equiv e^{i\omega_0 t}\int_{t_0}^t dt'\,e^{-i\omega_0 t'}F(t') \ .
\end{equation}
It follows from the expressions for $x$ and $\dot{x}$ in terms of $\zeta$ that
the total energy per unit mass of the oscillator as a function of time is given by
\begin{equation}
E(t) = \frac{1}{2}\left|\zeta(t)\right|^2 \ .
\end{equation}

Let the external force per unit mass now be of the form
\begin{equation*}
F(t) = F_0(t) \cos\left[\phi(t)\right] \ ,
\end{equation*}
with the amplitude $F_0$ and frequency $\dot{\phi}$ being slowly varying
functions of time, and $\ddot{\phi}>0$.
Resonance occurs when the relative phase of the driver and the oscillator
becomes stationary. This gives us the condition $\dot{\phi}(t) = \omega_0$.
We assume that there is only one passage through resonance,
and since there will be little average energy transfer away from
resonance, we need only consider the form of the forcing function
near resonance.
Let $t_R$ be the time when the resonance condition is satisfied,
and expand the driver in a Taylor series around this point:
\begin{equation}
F(t_R + \tau) \simeq F_0(t_R) \cos\left[\phi(t_R) + \omega_0\tau + \ddot{\phi}(t_R)\frac{\tau^2}{2}\right] \ .
\end{equation}
(Since the amplitude varies slowly with time, to lowest order, we can take the amplitude
as constant through the resonance.)
With the definitions
\begin{equation*}
F_R \equiv F_0(t_R) \ , \quad \phi_R \equiv \phi(t_R)\ , \quad \alpha \equiv \frac{\ddot{\phi}(t_R)}{\omega_0^2} \ ,
\end{equation*}
this becomes
\begin{equation} \label{eq:sho_acc}
F(t_R + \tau) \simeq F_R \cos\left(\phi_R + \omega_0\tau + \alpha\frac{\omega_0^2\tau^2}{2}\right) \ .
\end{equation}
The parameter $\alpha$ has the physical interpretation of being
a measure of the fractional change in frequency over a
characteristic period of oscillation.
The requirement that the frequency of the driver is varying slowly
therefore implies $\alpha\ll 1$. In other words, the driver can
be considered harmonic with a well-defined frequency over several
periods of the oscillator. We can also view $\alpha$ as a measure
of the phase ``drift''--i.e.~a measure of how fast the driver
accumulates additional phase. With this interpretation, it
is easy to see that the time spent near resonance is given
by $(\alpha\omega_0^2)^{-1/2}$, approximately.

The problem of solving for the motion of the oscillator is 
now mathematically similar to the theory of Fresnel diffraction,
and the solution can be expressed in terms of Fresnel integrals.
We relegate the detailed calculation to Appendix~\ref{APP:SHO}
and simply summarise the result:
\begin{equation*}
\begin{split}
\zeta_1 = \frac{F_R}{2}e^{i\omega_0 \tau} \Bigg\{ &I_1(\tau) \exp\left(i\phi_R\right) \\ &+ I_2(\tau) \exp\left[-i\left(\phi_R - \frac{2}{\alpha}\right)\right] \Bigg\} \ ,
\end{split}
\end{equation*} 
where $I_1$ and $I_2$ are given by expressions involving Fresnel integrals.
The time-averaged energy per unit mass changes asymptotically by
\begin{equation} \label{eq:sho_delta_E}
\Delta E = \frac{\pi F_R^2}{4\alpha\omega_0^2}\left( 1 + 2\sqrt{\frac{E_0}{\pi F_R^2/4\alpha\omega_0^2}}\cos\psi\right) \ ,
\end{equation}
where $E_0=\left|\zeta_0\right|^2/2$ is the initial energy,
and $\psi$ is an initial phase.
Qualitatively, the velocity is in quadrature with the force
well away from resonance, but the relative phase of the two becomes
approximately stationary near resonance for a time interval 
$\sim(\alpha\omega_0^2)^{-1/2}$, and there 
is a velocity change $\sim F_R (\alpha\omega_0^2)^{-1/2}$.
The presence of the $\psi$ dependent term reflects the fact that the oscillator
may gain or lose energy, depending upon its initial energy and
the relative phasing with the driver near resonance.
If we perform an ensemble average over initial phases
(assuming a uniform distribution),
we find that the average energy transfer is given by
\begin{equation}
\varepsilon\equiv\langle\Delta E\rangle = \frac{\pi F_R^2}{4\alpha\omega_0^2} \ .
\end{equation}
From (\ref{eq:sho_delta_E}) it is then obvious that, for
$E_0\ll \varepsilon$, the initial
phase is unimportant and the actual energy transfer will
be very close to the average. The possibility of
negative energy transfer only exists when
\begin{equation*}
E_0 > \frac{\varepsilon}{4 C^2} \ ,
\end{equation*}
where $C\equiv\cos\psi$.
Note that, since the energy of the oscillator cannot be
negative, it must be true that $\Delta E \ge -E_0$.
It can be shown that (\ref{eq:sho_delta_E}) complies
with this constraint.

Simple, linear damping is conventionally treated by adding a term
$2\gamma\dot{x}$ to the left side of (\ref{eq:sho}).
When $\gamma\ll(\alpha\omega_0^2)^{1/2}$, the development of the oscillation
will be  uninfluenced by damping, although the energy of the oscillation 
will be converted steadily into heat. However, when the damping is effective
on the time-scale of energy transfer, the amplitude of the oscillation will be
reduced. Nonetheless, it can be shown that the energy that appears ultimately
as heat is still given by (\ref{eq:sho_delta_E}), independent of $\gamma$, as
long as $\gamma\ll\omega_0$ \citep[see, for example, ][]{lan69}.

\subsection{White dwarf oscillations} \label{SS:WDO}
In this paper, we confine our attention to homogeneous, non-rotating 
white dwarfs where the pressure is contributed solely by cold, degenerate
electrons. Thermal corrections, Coulomb effects, as well as compositional
discontinuities are ignored. The relevant equations of stellar structure are
described in \citet{kip90}. We consider three cases with masses 
$0.6$, $1.0$, $1.4$~M$_\odot$ for $\mu_e=2$. Some relevant properties are given in
Table \ref{tab:wd_models}.

\begin{table*}
\begin{center}
\begin{tabular}{cccccc} \hline
Mass&Radius&$\omega_{f2}$&$M_{f2}$&$\eta_{3,f2}(R_{\ast})/\eta_{1,f2}(R_{\ast})$ & $\Theta_{f2}$ \\
(M$_{\odot}$)&($10^8$~cm)&($\omega_{\ast}$)&($10^{-2}\ M_{\ast}$)&  & \\ \hline\hline
$0.6$ & $8.83$ & $1.53$ & $2.05$ & $-0.169$ & $9.6\times 10^7$ \\
$1.0$ & $5.71$ & $1.65$ & $1.28$ & $-0.124$ & $8.0\times 10^6$ \\
$1.4$ & $1.98$ & $1.97$ & $0.25$ & $-0.0412$ & $1.5\times 10^5$ \\ \hline

\end{tabular}
\caption{Homogeneous, cold white dwarf models with $\mu_e=2$, and properties of
their quadrupolar \emph{f}-modes. \label{tab:wd_models}}
\end{center}
\end{table*}

The linear theory of normal modes for a cold white dwarf has been given by several 
authors \citep[e.g.][]{kip90}. We briefly review it to establish notation and to
resolve some ambiguities. It is convenient to use Eulerian perturbations and to
choose four independent variables which we write as
$\boldeta(r)Y_{\ell m}^{\sigma}(\theta,\varphi)e^{i\omega t}$, where 
\begin{equation}
\label{eq:eta}
\boldeta=\left(\frac{\xi_r}{r},\frac{\delta P+\rho\delta\Psi}{\rho g r},\frac{\delta\Psi}{g r},
\frac{\delta\Psi'}{g}\right) \ ,
\end{equation}
$\Psi$ is the Newtonian potential, $g\equiv\Psi'$ is the gravitational
acceleration, primes denote differentiation with respect to $r$, and
$Y_{\ell m}^{\sigma}$ are the real spherical harmonics defined for $m\ge 0$ by
\begin{align*}
Y_{\ell m}^{e} &\equiv \left\{\begin{array}{lll}Y_{\ell 0} & , \ m=0 \\ \left( Y_{\ell m} + Y_{\ell m}^* \right)/\sqrt{2} & , \ m>0 \end{array}\right. \\
Y_{\ell m}^{o} &\equiv \left\{\begin{array}{lll}0 & , \ m=0 \\ \left( Y_{\ell m} - Y_{\ell m}^* \right)/i\sqrt{2} & , \ m>0  \ .\end{array}\right.
\end{align*}
Our choice to use real rather than complex spherical harmonics allows us to avoid
complications that would arise from the expansion of a real vector field in terms
of complex basis vectors.
Note that the Cowling approximation, which ignores the potential variation, is commonly 
used to describe stellar pulsations, but it is inadequate for our calculations.
We denote the spatial displacement field for a mode by
$\bxi_j({\bld{r}})$,\footnote{A word on notation: we use the the index $j$ as shorthand for the complete set
of labels required to specify a mode uniquely. In places where we need to
distinguish between even ($\propto\cos(m\varphi)$) and odd ($\propto\sin(m\varphi)$) modes,
we use the additional labels $e$ and $o$ for even and odd, respectively. In such places, $j$ should be
understood to mean the set of all other labels. Occasionally, we use the
index $\sigma$ which can take on the values $e$ or $o$.} the components of which are given by
\begin{equation}
\bxi_j(\bld{r}) = \left( r\eta_{1j} \bld{\hat{r}} + \frac{g}{\omega_j^2}\eta_{2j} \nabla\right) Y_{\ell m}^{\sigma}(\theta,\varphi) \ .
\end{equation}

With the above definitions, the continuity, Euler, entropy, and Poisson equations
yield four first-order ordinary differential equations \citep[e.g.][]{kip90}.
There are two boundary conditions at the centre of the star,
\begin{equation*}
\eta_1 = \frac{g \eta_2}{r \omega^2} \ , \quad \eta_3 = \frac{\eta_4}{\ell} \ ,
\end{equation*}
and two more at the surface,
\begin{equation*}
\eta_1 = \eta_2 - \eta_3 \ , \quad \eta_3 = -\frac{\eta_4}{\ell+1} \ .
\end{equation*}
With a choice of normalisation, the system of equations for $\boldeta$ can be
solved numerically to get the normal mode frequencies and eigenfunctions.
The usual choice is to normalise the eigenfunctions by mass so that
\begin{equation*}
\int d^3x \, \rho_0 \bxi_j\mathbf{\cdot}\bxi_j' = \delta_{j j'}
\end{equation*}
\citep[see, for example,][]{pre77}, where $\rho_0$ is the unperturbed mass density.
However, we make the non-standard choice $\eta_1(R_{\ast})=1$, where
$R_{\ast}$ is the radius of the white dwarf. This
ensures that the coefficient of a mode in an
eigenfunction expansion is a measure of the relative surface displacement and,
consequently, a measure of the mode non-linearity.
With our normalisation convention, we define the mode ``mass'' by
\begin{equation}
M_j \equiv \frac{1}{R_{\ast}^2} \int d^3x \, \rho_0 |\bxi_j|^2 \ .
\end{equation}
It is convenient for us to work in terms of dimensionless
quantities whenever possible, and towards that end
we shall make use of the definitions
\begin{align*}
&\omega_{\ast} \equiv \sqrt{\frac{G M_{\ast}}{R_{\ast}^3}} \ , & \sigma_j &\equiv \frac{\omega_j}{\omega_{\ast}} \ , \\
& \beta_{\ast} \equiv  \frac{1}{c}\sqrt{\frac{G M_{\ast}}{R_{\ast}}} \ , & E_{\ast} &\equiv \frac{G M_{\ast}^2}{R_{\ast}} \ ,
\end{align*}
where $M_{\ast}$ is the mass of the white dwarf. The physical significance
of these definitions is that, to within factors of order unity,
$\omega_*$ is a typical mode frequency scale,
$\beta_*$ is the surface escape speed, and $E_*$ is the gravitational binding
energy of the white dwarf.

\begin{figure}
\includegraphics[width=\columnwidth]{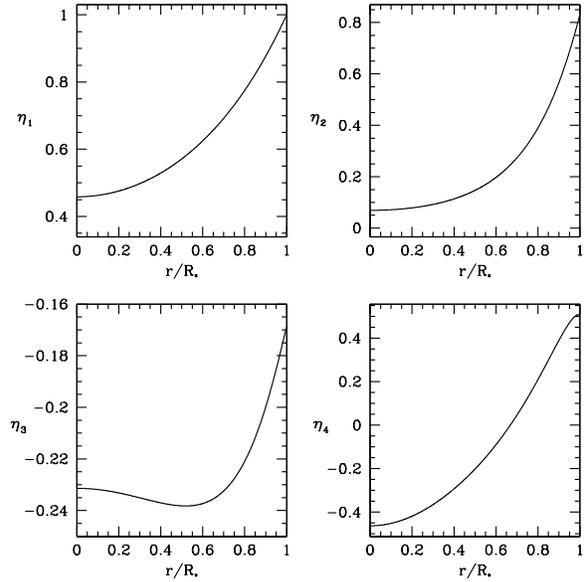}
\caption{\label{fig:wd_efn} Radial eigenfunctions of quadrupolar \emph{f}-modes
for the $0.6\,M_{\odot}$ model from Table~\ref{tab:wd_models}.}
\end{figure}

The most important modes for our purpose are the quadrupolar \emph{f}-modes. For a
non-rotating star, the five \emph{f}-modes with $\ell=2$ are degenerate
in frequency. The eigenfrequencies for our three white dwarf
models are given in Table~\ref{tab:wd_models}. The radial eigenfunctions for
the $0.6\,M_{\odot}$ model are displayed in Figure~\ref{fig:wd_efn}. The
eigenfunctions for the other white dwarf models are qualitatively similar.

\subsection{Gravitational radiation}
We adopt a Newtonian approach to gravitational
radiation reaction in the two-body problem, neglecting all finite-size
effects. Namely, we treat the problem as essentially Keplerian with
prescribed corrections to the orbital equations. For non-relativistic orbits
($v\lesssim 0.2c$), the secular corrections due to
gravitational radiation are provided to a fair approximation by the
orbit-averaged expressions
\begin{align}
\frac{d E_{\rm orb}}{d t} &= -\frac{32}{5} E_{\ast}\omega_{\ast}\frac{q^2}{(1+q)^{2/3}} \beta_{\ast}^5 \left(\frac{n}{\omega_{\ast}}\right)^{10/3} \mathcal{F}_1(e) \ , \\
\frac{d L_{\rm orb}}{d t} &= -\frac{32}{5} E_{\ast} \frac{q^2}{(1+q)^{2/3}} \beta_{\ast}^5 \left(\frac{n}{\omega_{\ast}}\right)^{7/3} \mathcal{F}_2(e)
\end{align}
\citep{pet64}, where $E_{\rm orb}$ and $L_{\rm orb}$ are the orbital energy
and angular momentum, $q$ is the ratio of the companion mass to the
white dwarf mass, $n$ is the Keplerian orbital frequency,
$e$ is the orbital eccentricity, and
\begin{align*}
\mathcal{F}_1(e) &\equiv \frac{1}{(1-e^2)^{7/2}}\left(1+\frac{73}{24}e^2+\frac{37}{96}e^4\right) \ , \\
\mathcal{F}_2(e) &\equiv \frac{1}{(1-e^2)^2}\left(1+\frac{7}{8}e^2\right) \ .
\end{align*}
We can re-express the orbital evolution in terms of changes in the
orbital frequency and eccentricity:
\begin{align}
\label{eq:gr_n}
\frac{d n}{d t} &= \frac{96}{5} \omega_{\ast}^2 \frac{q}{(1+q)^{1/3}}\beta_{\ast}^5 \left(\frac{n}{\omega_{\ast}}\right)^{11/3} \mathcal{F}_1(e) \ , \\
\label{eq:gr_e}
\frac{d e}{d t} &= -\frac{304}{15} \omega_{\ast} \frac{q}{(1+q)^{1/3}}\beta_{\ast}^5 \left(\frac{n}{\omega_{\ast}}\right)^{8/3} \mathcal{F}_3(e) \ ,
\end{align}
where
\begin{equation*}
\mathcal{F}_3(e) \equiv \frac{e}{(1-e^2)^{5/2}}\left(1+\frac{121}{304}e^2\right) \ .
\end{equation*}
If gravitational radiation is the only mechanism for orbital evolution, then
it follows from these equations that
\begin{equation} \label{eq:gr_edot_from_ndot}
\dot{e} = -\mathcal{G}(e)\frac{\dot{n}}{n} \ ,
\end{equation}
where
\begin{equation*}
\mathcal{G}(e) \equiv \frac{19}{18}\frac{\mathcal{F}_3(e)}{\mathcal{F}_1(e)} \ .
\end{equation*}
Finally, we can integrate the above equation to get
\begin{equation} \label{eq:gr_n_of_e}
\frac{n(e)}{n(0.54101)} = \frac{(1-e^2)^{3/2}}{e^{18/19}}\left(1+\frac{121}{304}e^2\right)^{-1305/2299} \ .
\end{equation}
As the orbit shrinks, it circularises, eventually according to $e\propto n^{-1}$,
approximately.

For a more accurate treatment of gravitational radiation (especially for high
eccentricities), and for the inclusion of other general relativistic effects,
corrections to the orbital acceleration can be added
directly to the equations of motion. Detailed derivations and discussions of these
corrections can be found in the literature \citep[e.g.][]{iye95}, and we shall not
reproduce them here.

It should be noted that it is not necessary to worry about relativistic apsidal
precession as it will only affect neglected higher-order terms.

\subsection{Equations of motion}
Neglecting general-relativistic effects, the equations of motion for a
white dwarf-compact object binary can be arrived at via several approaches.
A straightforward, Newtonian ``push-pull'' approach to the excitation
of tides and the resulting perturbations to the orbit is perhaps the most
intuitive. However, it is not transparently self-consistent in terms of conserving
energy and angular momentum. A variational approach has the advantage of
maintaining self-consistency explicitly, and also has the potential for exposing
underlying symmetries \citep{gin80,rat03}.

We expand the physical displacement, $\bxi$, of fluid elements within the
white dwarf in terms of normal modes,
\begin{equation}
\bxi(\bld{r},t) = \sum_j A_j(t) \bxi_j(\bld{r}) \ .
\end{equation}
Note that, as mentioned in \S\ref{SS:WDO}, we are not using the conventional
normalisation for the normal modes. With our normalisation, the $\bxi_j$ have
dimensions of length and the mode displacements $A_j$ are dimensionless.
The energy and angular momentum in each even-odd mode pair can be written in terms
of their displacements $A_{j\sigma}(t)$ as
\begin{equation*}
E_j = \frac{1}{2}M_j R_{\ast}^2 \left[ \dot{A}_{je}^2 + \dot{A}_{jo}^2 + \omega_{j}^2 \left(A_{je}^2 + A_{jo}^2 \right) \right]
\end{equation*}
and
\begin{equation*}
L_j = m M_j R_{\ast}^2\left(\dot{A}_{je} A_{jo} - \dot{A}_{jo} A_{je}\right)
\end{equation*}
\citep{rat03}, respectively. For an isolated mode (i.e.~one not being excited), the
following simple relation also holds:
\begin{equation*}
L_j = m\frac{E_j}{\omega_j} \ .
\end{equation*}
The conserved energy and angular momentum for the entire system are given by
\begin{equation} \label{eq:cons_en}
\begin{split}
E =\ &\frac{1}{2}\mu\left(\dot{R}^2 + R^2\dot{u}^2\right) - \frac{G M_0 M_{\ast}}{R} \\ &+ \sum_j E_j - \sum_j A_j f_j(\bld{R})
\end{split}
\end{equation}
and
\begin{equation}
L = \mu R^2\dot{u} + \sum_j L_j \ ,
\end{equation}
where $R$ is the orbital separation, $u$ is the angular coordinate in the plane
of the orbit, $M_0$ is the companion mass, $\mu$ is the reduced mass, and
\begin{equation} \label{eq:overlap}
f_j(\bld{R}) = G M_0\int d^3r\, \rho_0\bxi_j\cdot\nabla\left(\frac{1}{|\bld{r}-\bld{R}|}\right)
\end{equation}
\citep{rat03}.

We can use (\ref{eq:cons_en}) as the Hamiltonian for the system, and with
the definition $x_j\equiv R_{\ast}A_j$, it takes the simple form
\begin{equation}\label{eq:hamiltonian}
\begin{split}
H &= \frac{p_R^2}{2\mu} + \frac{p_u^2}{2\mu R^2} - \frac{G M_0 M_*}{R} \\ &+ \sum_j \left( \frac{p_j^2}{2 M_j} + \frac{1}{2}M_j\omega_j^2 x_j^2 \right) - \sum_j x_j f_j \  ,
\end{split}
\end{equation}
where $p_R=\mu\dot{R}$, $p_u=\mu R^2\dot{u}$, and $p_j=M_j\dot{x}_j$ are the momenta
conjugate to $R$, $u$, and $x_j$, respectively. We therefore see that the Hamiltonian
is comprised of three pieces: the Keplerian terms for the orbit, a sum of independent
harmonic oscillators for the normal modes, and a sum of terms of the form $x_j f_j$
which couple the modes and the orbit. The overlap integral $f_j$ therefore plays
a dual role as a driving function for the excitation of tides, and in the
disturbing function for the orbit. This is not surprising since the system is
conservative. Hence, any energy and angular momentum transferred to tides must
be extracted necessarily from the orbit.

From Hamilton's equations for (\ref{eq:hamiltonian}), we obtain the following
equations of motion:
\begin{gather}
\ddot{x}_j + \omega_j^2 x_j = \frac{f_j}{M_j} \label{eq:mode_ex}  \ , \\
\dot{p}_R = \frac{p_u^2}{\mu R^3} - \frac{G M_0 M_*}{R^2} + x_j\frac{\partial f_j}{\partial R} \ , \\
\dot{p}_u = x_j\frac{\partial f_j}{\partial u} \ .
\end{gather}
The terms involving the derivatives of $f_j$ give the perturbation of the orbit
due to the excitation of tides, and we therefore refer to them as the back
reaction terms. The overlap integral $f_j$ can also be written as
\begin{equation}
f_j(\bld{R}) = -q\eta_jM_{\ast}\omega_{\ast}^2 R_{\ast}\left(\frac{R_{\ast}}{R}\right)^{\ell+1} \left\{\begin{array}{l}\cos (m u) \\ \sin (m u) \end{array}\right. \ ,
\end{equation}
where
\begin{equation*}
\eta_j \equiv \eta_{3j}(R_{\ast}) Y_{\ell m}^{e}\left(\frac{\pi}{2},0\right)
\end{equation*}
\citep{rat03}, and the bracket notation denotes that either $\cos(mu)$ or $\sin(mu)$
will be present. It will be useful for us to write $f_j$ in yet another way.
From the usual Keplerian relation between the orbital frequency $n$ and the
semi-major axis $a$, it follows that
\begin{equation*}
f_j = -\frac{q\eta_j M_{\ast}\omega_{\ast}^2 R_{\ast}}{(1+q)^{(\ell+1)/3}} \left(\frac{n}{\omega_{\ast}}\right)^{2(\ell+1)/3}\left(\frac{a}{R}\right)^{\ell+1} \left\{\begin{array}{l}\cos(m u)\\\sin(m u)\end{array}\right. \ .
\end{equation*}
We now make use of the Fourier expansion
\begin{equation*}
\left(\frac{R}{a}\right)^p \exp(imv) = \sum_{k= -\infty}^{\infty} X_k^{p,m}(e) \exp(ikl) \ ,
\end{equation*}
where $v$ is the true anomaly, $l$ is the mean anomaly
(not to be confused with $\ell$), and the Fourier coefficients $X_k^{p,m}$ (called
Hansen coefficients; see Appendix~\ref{APP:HC}) are real functions of
the eccentricity. Noting that $u = v + \varpi$, where $\varpi$ is the
longitude of periapse, we have
\begin{equation} \label{eq:hansen_coeff}
\begin{split}
\left(\frac{a}{R}\right)^{\ell+1} \left\{ \begin{array}{l} \cos(m u) \\ \sin(m u) \end{array} \right. = \sum_{k= -\infty}^{\infty} &X_k^{-(\ell+1),m}(e) \\ &\times\left\{ \begin{array}{l} \cos(kl+m\varpi) \\ \sin(kl+m\varpi) \end{array} \right. .
\end{split}
\end{equation}
The overlap integral $f_j$ is therefore given by
\begin{equation}
f_j = \sum_{k=0}^{\infty} f_{jk}
\end{equation}
where
\begin{equation}\label{eq:overlap_jk}
\begin{split}
f_{jk} &= -\frac{q\eta_j M_{\ast}\omega_{\ast}^2 R_{\ast}}{(1+q)^{(\ell+1)/3}} \left(\frac{n}{\omega_{\ast}}\right)^{2(\ell+1)/3} \\ &\times\left\{ \begin{array}{l} \left[X_{jk}^+\cos(kl+m\varpi) + X_{jk}^-\cos(kl-m\varpi)\right] \\ \left[X_{jk}^+\sin(kl+m\varpi) - X_{jk}^-\sin(kl-m\varpi)\right] \end{array} \right. \ ,
\end{split}
\end{equation}
and we have used the shorthand $ X_{jk}^{\pm} \equiv X_{\pm k}^{-(\ell+1),m}$,
for economy of notation. It should be understood in the expression for $f_{jk}$ that,
for $k=0$, only the $X_{jk}^+$ terms are present. For $k>0$, the $X_{jk}^{\pm}$
terms can be combined using trigonometric identities. However, it is simpler to note
that, since $X_k^{p,m} \propto e^{\left|k-m\right|}$, to lowest order in eccentricity,
the $X_{jk}^-$ terms will be suppressed by $2m$ powers of eccentricity relative to the
$X_{jk}^+$ terms. Therefore, for low to moderate eccentricities
($\lesssim 0.6$) and $m>0$, the $X_{jk}^-$ terms can be neglected to a
good approximation. For the case $m=0$, the $X_{jk}^{\pm}$ terms are
identical. Hence, in all that follows, for $m=0$ one only needs
to make the change $X_{jk}^+\rightarrow 2 X_{jk}^+$.

From the preceding discussion, we know that the driving function $f_j$ for the
excitation of a particular mode is an infinite sum of $f_{jk}$ terms.
The phases that appear in the expression (\ref{eq:overlap_jk}) for $f_{jk}$ are
all of the form $kl\pm m\varpi$. Thus, there exists the possibility of resonance
whenever the relative phase of the mode and one of these terms is stationary:
$\dot{w}_j = k\dot{l} \pm m\dot{\varpi}$, where $w_j$ is the phase of the mode.
As mentioned previously, the $kl-m\varpi$ terms will be suppressed by $2m$
powers of eccentricity relative to the $kl+m\varpi$ terms. Thus, the dominant
resonances will occur for $\dot{w}_j  = k\dot{l} + m\dot{\varpi}$.
It might be thought that the above condition is equivalent to $\omega_j = kn$, but,
in general, this is not the case. As the evolution of the orbit is dependent
upon the tides via the back reaction terms in the equations of motion, there are
complicated, non-linear dependencies implicit in each of the variables in the
resonance condition. However, since we expect the orbital corrections to be
relatively small, it should be true that, at resonance, $\omega_j\simeq k n$.

\section{Physical Considerations}
\subsection{The tidal limit}
Clearly, our formalism for treating the evolution of a WDCO binary
as a dynamical interaction between the orbit and the
tides is only valid if the white dwarf is not tidally disrupted. In
other words, we require that the white dwarf does not fill its Roche
lobe. This requirement constrains the harmonics of the orbital
frequency that a given mode can interact resonantly with. To quantify
the constraint, we use the following approximation to the radius of the Roche
lobe:
\begin{equation*}
\frac{r_{\rm R}}{R} = \frac{0.49 q^{-2/3}}{0.6 q^{-2/3} + \ln(1+q^{-1/3})}
\end{equation*}
\citep{egg83}. It then follows that we require
\begin{equation}
k \gtrsim \frac{2.92 \sigma_j}{(1-e)^{3/2}} \frac{\left[0.6 + q^{2/3}\ln(1+q^{-1/3})\right]^{3/2}}{(1+q)^{1/2}} \ ,
\end{equation}
where we have made use of the facts that the orbital separation at periapse is
$a(1-e)$, and that $\omega_j\simeq kn$ at resonance. Note that we have
implicitly assumed that the companion is more compact than the white
dwarf, and hence is not disrupted. This is certainly true when the
companion is a neutron star or a black hole.
However, for the white dwarf-white dwarf case, the actual
constraint is provided by the star that is disrupted first, which
may be the companion.

It should also be mentioned that the above approximation for the radius
of the Roche lobe is for circular, synchronous orbits. A more general
treatment of the Roche problem may modify the tidal disruption regime.
This is a possibility for future investigation.

\subsection{Importance of the $\ell=m=2$ \emph{f}-mode}
The lowest $\ell$ modes that can be excited tidally are $\ell=2$. Modes with higher
values of $\ell$ will have smaller overlap integrals, since $f_j\propto R^{-(\ell+1)}$.
We may therefore infer that the primary modes that are excited outside
the Roche limit are the $\ell=2$ modes. It is also the case that, with our choice
of coordinates, the $m=1$ modes will not be excited. This is easily seen by
remembering that $\eta_j\propto P_{\ell}^m(0)$, and
\begin{equation*}
P_{\ell}^m(0) = \left\{\begin{array}{cl} (-1)^{(\ell-m)/2} \dfrac{(\ell+m-1)!!}{(\ell-m)!!} \ , & \ell+m\ \mathrm{even} \\ 0 \ , & \ell+m\ \mathrm{odd} \end{array}\right.
\end{equation*}
\citep[see, for example,][]{arf95}.
Therefore, the only $\ell=2$ modes that are excited have $m=0,2$. Furthermore, since
$X_k^{p,m}\propto e^{\left|k-m\right|}$, the $m=0$ modes will be suppressed by
two powers of eccentricity relative to the $m=2$ modes. Hence, we deduce that
the dominant modes for low to moderate eccentricities will have
$\ell=m=2$. Also, since the \emph{p}-mode frequencies increase monotonically with
the radial order, we can access (before tidal disruption) the lowest
harmonic resonances for the modes with lowest radial order--the \emph{f}-modes.

Putting together the above considerations, we conclude that the mode excited with
the largest amplitude in a cold white dwarf will be the $\ell=m=2$ \emph{f}-mode.
Note that in a warm star, \emph{g}-modes can also be excited. These will have lower
frequencies than the \emph{f}-modes, and their frequencies will decrease monotonically
with the radial order. However, the structure of \emph{g}-modes is sensitive
to assumptions about the stellar model. If the modes are confined to surface layers,
then the overlap integrals will be essentially zero, and the modes will not be excited
tidally.

\subsection{Mode damping}
The formalism that we have presented in \S\ref{S:GF} does not include any mode damping.
In a realistic scenario, white dwarf oscillations will damp out over sufficiently long
periods of time. While we shall mention some possible mechanisms through which this might
occur, we make no attempt to provide an exhaustive analysis as there is an extensive
literature that exists for this problem.

Some possible mechanisms that have been considered for the damping of nonradial white dwarf
oscillations include gravitational radiation, neutrino losses due to pycnonuclear reactions,
and radiative heat leakage \citep{osa73}. The relative importance of each mechanism depends
on the type of mode under consideration, but it was demonstrated by \citet{osa73} that
the dominant damping mechanism for quadrupolar \emph{f}- and \emph{p}-modes,
in the linear regime, is gravitational radiation. However, their calculation contains a
numerical error. We present a corrected derivation in Appendix~\ref{APP:DOMUGR}.

Another possible mechanism for the damping of modes with large amplitudes
is by non-linear coupling to other modes. This has been explored extensively
in various contexts \citep[e.g.][]{dzi82,kum96,wu01}, and it has been shown that
non-linear mode interactions can be important amplitude limiting effects.
In this paper, we ignore this complication because it is, in fact, one of
our goals to study whether such non-linear amplitudes can be excited by
passage through a sequence of tidal resonances in a WDCO binary.

In stars with compositional discontinuities or solid interiors, turbulence
may be excited at boundaries, which can lead to additional dissipation.

\subsection{Time-scales}
For the long term evolution of a WDCO binary, there are several
time-scales of interest to us.
The first of these is the gravitational radiation inspiral time, which,
for a circular orbit, is given by
\begin{equation*}
T_{\rm GR} = \frac{5}{256\omega_{\ast}}\frac{(1+q)^{1/3}}{q} \beta_{\ast}^{-5} \left(\frac{n}{\omega_{\ast}}\right)^{-8/3} \ .
\end{equation*}
\citep{pet64}. For an eccentric orbit, the inspiral time has to be calculated
numerically. In general, for an eccentric orbit with a given period, this time
is shorter by up to a factor of $1000$ for eccentricities up to $0.9$, relative
to the circular orbit time. For eccentricities $\lesssim0.5$, however, the
circular orbit inspiral time is a fair approximation.

The second relevant time-scale is the mode damping time. In general, the damping times for
quadrupolar \emph{f}-modes depend upon the white dwarf mass. Assuming gravitational
radiation as the mechanism, the damping time (as derived in Appendix~\ref{APP:DOMUGR})
is given by
\begin{equation*}
T_j = \frac{6\pi}{\omega_{\ast}}\beta_{\ast}^{-5}
\eta_{3j}^{-2}(R_{\ast})\left(\frac{M_j}{M_{\ast}}\right)
\sigma_j^{-4} \ .
\end{equation*}
For our $0.6$~$M_\odot$ and $1.0$~$M_\odot$ models, this gives $\sim 3000$ and
$\sim 100$ years, respectively. Note that these are necessarily underestimates
since our cold white dwarf models are highly centrally condensed. In contrast, the
damping times for ``moderately realistic'' $0.4$~$M_\odot$ and $1.0$~$M_\odot$ models
used by \citet{osa73} are about $2.8\times 10^5$ and $500$ years,
respectively. The damping times are therefore quite sensitive to the
stellar model.

Finally, the third time-scale of interest is the white dwarf cooling time. A rough estimate
for this is provided by
\begin{equation*}
T_{\rm cool} = \frac{4.7\times 10^7 \ \rm years}{A} \left(\frac{M_\ast/M_\odot}{L_\ast/L_\odot}\right)^{5/7}
\end{equation*}
\citep{kip90}, where $A$ is the atomic mass, and $L_\ast$ is the white dwarf luminosity.
For typical parameters, this gives a cooling time of $\sim10^9$ years, which is much longer
than any other relevant time-scale. We can therefore ignore the thermal
evolution of the white dwarf.

In order for mode damping via gravitational radiation to be physically
unimportant during the long-term evolution of a WDCO system, it is
necessary that $T_j>T_{\rm GR}$. In other words, we require that the
damping between resonances is negligible during the gravitational
inspiral. This gives us the following constraint on the harmonics that
we can consider for a particular mode:
\begin{equation}
k \lesssim \left[\frac{1536\pi}{5}\frac{q}{(1+q)^{1/3}}\eta_{3j}^{-2}(R_{\ast})\left(\frac{M_j}{M_{\ast}}\right)\sigma_j^{-4/3}\right]^{3/8} \ ,
\end{equation}
where we have used the expressions for $T_j$ and $T_{\rm GR}$ given above,
and have also made use of $\omega_j\simeq kn$ at resonance.
For our $0.6$~$M_{\odot}$ white dwarf model and mass ratios greater than
a few, this constraint evaluates to
\begin{equation*}
k \lesssim 11 \left(\frac{M_0}{M_{\odot}}\right)^{1/4} \ .
\end{equation*}
Note that, for moderate to high eccentricities, this is overly restrictive,
and the actual limit obtained from an evaluation of the inspiral time for
eccentric orbits is higher.

\section{Resonant Energy Transfer}\label{SEC:RET}
Let us now consider a mode being excited resonantly on a white dwarf
in an eccentric orbit around a compact companion.
We shall neglect the back reaction terms in the equations of motion,
and hence the orbit can be taken to be Keplerian with corrections due
to gravitational radiation  (the validity of the no back reaction
approximation will be discussed in \S\ref{SSEC:ROV}).
We assume that we start exciting the mode resonantly at $t=0$, and limit
our analysis to the regime $\dot{n}t/n \ll 1$,
where $\dot{n}$ is given by (\ref{eq:gr_n}).
This is not particularly restrictive since the gravitational radiation time-scale
is typically much longer than the resonance time-scale. Finally, we shall also
assume low to moderate eccentricities ($\sim 0-0.5$), and hence neglect the
$X_{jk}^-$ terms in (\ref{eq:overlap_jk}).

With the above assumptions, we can expand the orbital elements and phases in
(\ref{eq:overlap_jk}) in Taylor series around resonance (retaining only the
zeroth-order term in the amplitude) to obtain
\begin{equation}\label{eq:overlap_jk2}
\begin{split}
f_{jk} = -\frac{q\eta_j M_{\ast}\omega_{\ast}^2 R_{\ast}}{(1+q)^{(\ell+1)/3}} &\left(\frac{n}{\omega_{\ast}}\right)^{2(\ell+1)/3} X_{jk}^+ \\ &\times\left\{ \begin{array}{l} \cos\left(\phi_{jk} + \omega_j t + k\dot{n}t^2\right) \\  \sin\left(\phi_{jk} + \omega_j t + k\dot{n}t^2\right) \end{array} \right. \ ,
\end{split}
\end{equation}
where $\phi_{jk}$ is an initial phase. We now note
that (\ref{eq:overlap_jk2}) is exactly of the form of (\ref{eq:sho_acc}),
with the identifications
\begin{align*}
F_R &= -\frac{q\eta_j\omega_*^2R_*}{(1+q)^{(\ell+1)/3}}\left(\frac{M_{\ast}}{M_j}\right)\left(\frac{n}{\omega_*}\right)^{2(\ell+1)/3}  X_{jk}^+ \ , \\
\omega_0 &= \omega_j, \quad \alpha\omega_0^2 = 2k\dot{n}
\end{align*}
(the division by $M_j$ in $F_R$ is necessary since it is $f_j/M_j$ that appears on
the right hand side of (\ref{eq:mode_ex})). We can therefore immediately write down
the resonant energy transfer:
\begin{equation*}
\begin{split}
\frac{\langle\Delta E_{jk}\rangle}{E_{\ast}} = \frac{5\pi}{768}&\frac{q\beta_{\ast}^{-5}\eta_j^2}{(1+q)^{(2\ell+1)/3}}\left(\frac{M_{\ast}}{M_j}\right) \\ &\times\left(\frac{n}{\omega_{\ast}}\right)^{(4\ell-7)/3}\frac{\left(X_{jk}^+\right)^2}{k\mathcal{F}_1} \ ,
\end{split}
\end{equation*}
where we have averaged over initial phases. Using the fact that
$\omega_j\simeq kn$ at resonance, we find
\begin{equation}\label{eq:nbr_ecc_en_tr}
\frac{\langle\Delta E_{jk}\rangle}{E_{\ast}} = \frac{q}{(1+q)^{(2\ell+1)/3}}\Theta_j \Xi_{jk}(e) \ ,
\end{equation}
where the parameter
\begin{equation}
\Theta_j \equiv \frac{5\pi}{768}\beta_{\ast}^{-5}\eta_j^2\left(\frac{M_{\ast}}{M_j}\right)\sigma_j^{(4\ell-7)/3}
\end{equation}
depends only upon the white dwarf model and the mode, and
\begin{equation}
\Xi_{jk}(e) \equiv \frac{k^{-4(\ell-1)/3}}{\mathcal{F}_1}\left(X_{jk}^+\right)^2
\end{equation}
contains all the dependence upon the eccentricity and the harmonic.
The values of the parameter $\Theta_j$ for our $0.6$~$M_\odot$,
$1.0$~$M_\odot$, and $1.4$~$M_\odot$ white dwarf models are given in
Table~\ref{tab:wd_models}.
We see that the energy transfer decreases monotonically (relative to the star's
binding energy) with the mass. As $\Xi_{jk}(e) \propto e^{2(k-m)}$,
to lowest order in eccentricity, the energy transfer is typically a
very sensitive function of the eccentricity. Also, for a circular orbit,
it is clear that only the fundamental resonance, $k=m$, exists (as would
be expected on physical grounds). We remind the reader that, for $m\ne 0$,
the energy transfer given by (\ref{eq:nbr_ecc_en_tr}) is for a particular
choice of even or odd mode. It should therefore be multiplied by a factor
of two to obtain the total energy transfer to the even-odd mode pair.

\section{Discussion}
\subsection{Regime of validity}\label{SSEC:ROV}
\begin{figure*}
\includegraphics[width=5.0in]{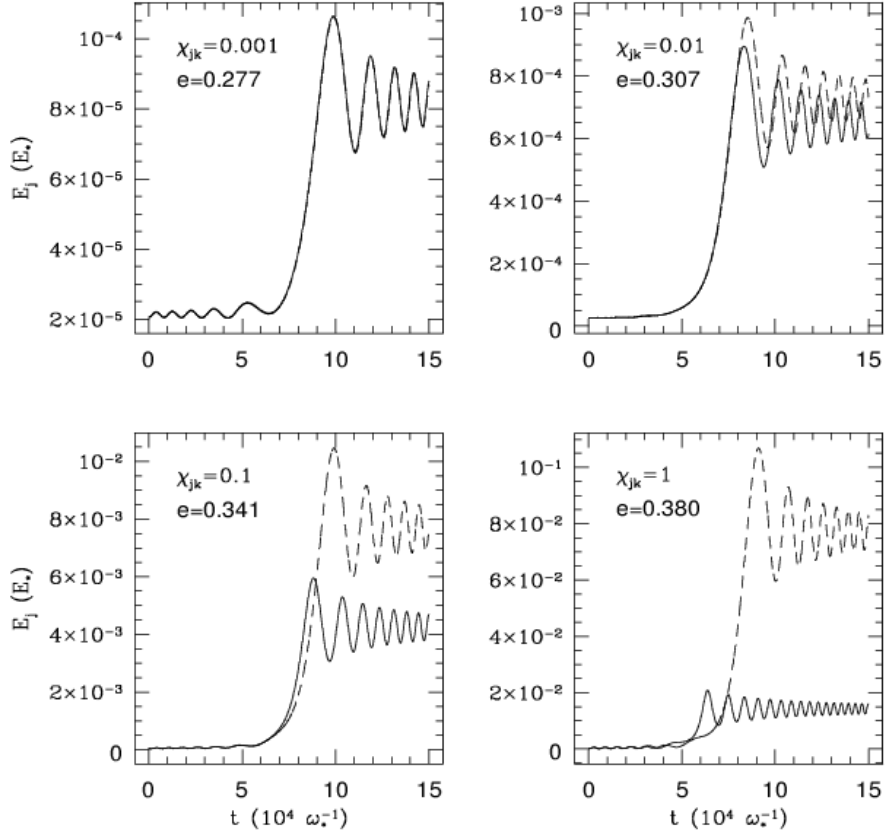}
\caption{\label{fig:br_nobr} The energy in the $\ell=m=2$ \emph{f}-mode
on a $0.6$~$M_\odot$ white dwarf is shown for a passage through the $k=15$ resonance
with different values of the parameter $\chi_{jk}$ obtained by varying the
eccentricity, and with $q=10,000$. In each plot, the dashed curve is the system
without back reaction, and the solid curve is the system with back reaction. The
curves have been smoothed to remove high-frequency components.}
\end{figure*}

We now consider in what regime, if any, the no back reaction approximation is
valid. Qualitatively, we expect back reaction to change the orbital frequency as a mode is
excited resonantly, which will tend to push the system away from resonance. Clearly, this will
modulate the energy transfer at some level. However, if the change in orbital frequency
is small compared to the resonance width, then we expect that the modulation of energy transfer
will not be significant. On the other hand, if the change in orbital frequency is comparable to
or larger than the resonance width, then back reaction will play a significant role. Another
way of saying this is that the modulation of the energy transfer by back reaction is a
second-order effect. Therefore, as long as the energy transfer is small enough, we are
justified in ignoring back reaction. We can quantify this criterion by defining a resonance
parameter
\begin{equation}
\chi_{jk} \equiv \frac{\Delta n_{jk}}{\Delta n^{\rm res}_k} \ ,
\end{equation}
where $\Delta n_{jk}$ is what the change in $k n$ would be if the energy given by
(\ref{eq:nbr_ecc_en_tr}) were to be taken out of the orbit, and $\Delta n^{\rm res}_k$ is
the resonance width. In general, we expect that for $\chi_{jk}\ll 1$ back
reaction will not play a significant role in modulating the energy
transfer, where as for $\chi_{jk}\gtrsim 1$ back reaction will be important.
Using the estimate $\Delta n^{\rm res}_k\approx (2k\dot{n})^{1/2}$, we
find
\begin{equation}
\begin{split}
\chi_{jk} = \sqrt{\frac{5}{3}}\frac{5\pi}{2048}&\frac{\beta_{\ast}^{-15/2}\eta_j^2\sigma_j^{(8\ell-23)/6}}{q^{1/2}(1+q)^{(4\ell-1)/6}}\left(\frac{M_{\ast}}{M_j}\right) \\ &\times\frac{k^{-(4\ell-7)/3}}{\mathcal{F}_1^{3/2}} \left(X_{jk}^+\right)^2 \ .
\end{split}
\end{equation}
Figure~\ref{fig:br_nobr} shows the numerical integration across a particular
resonance for various values of $\chi_{jk}$, both with and without back reaction.
The first qualitative feature that stands out is that the energy transfer with back
reaction tends to be smaller than that without back reaction. This is not surprising
since the system with back reaction is expected to spend less time near resonance.
Quantitatively, we see that for this particular resonance with $\chi_{jk}\lesssim 0.1$
we obtain nearly identical numerical results with and without back reaction, with
$\chi_{jk}\sim 0.1$ the results differ by a factor of order unity (about $2$),
and with $\chi_{jk}\sim 1$ the energy transfers differ by an order of magnitude.

\begin{figure*}
\begin{tabular}{cc}
\includegraphics[width=2.65in]{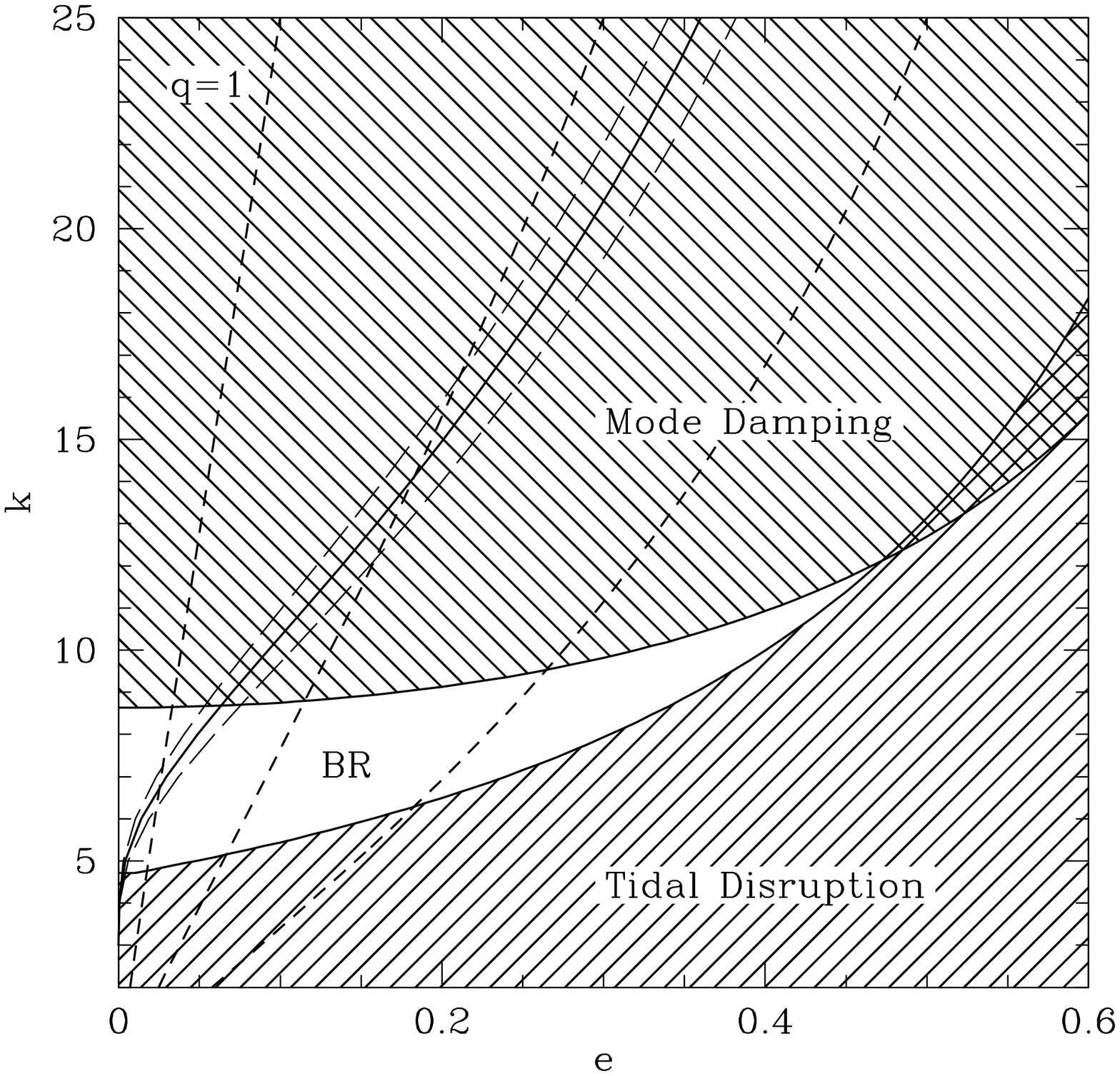} &
\includegraphics[width=2.65in]{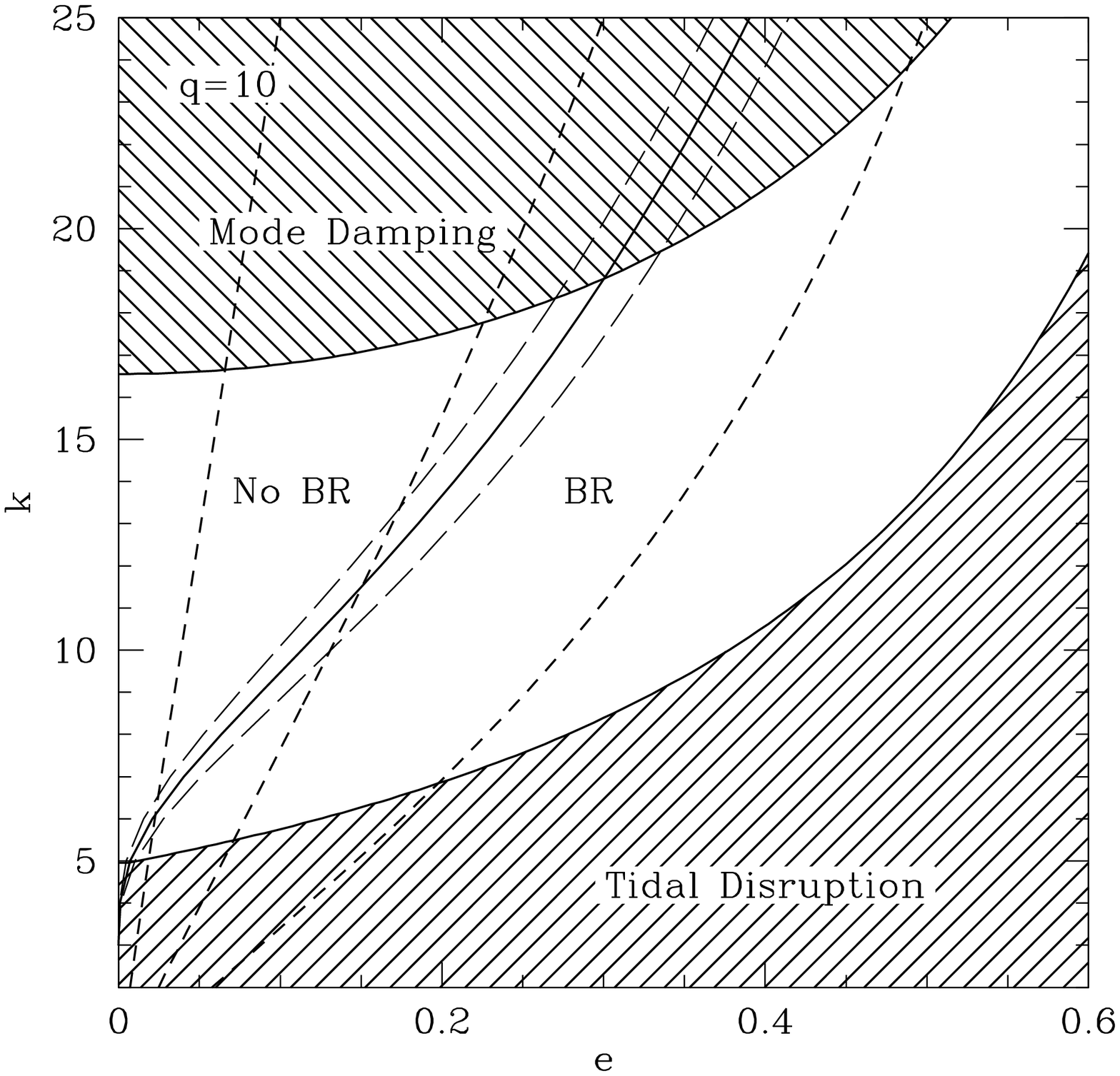} \\
\includegraphics[width=2.65in]{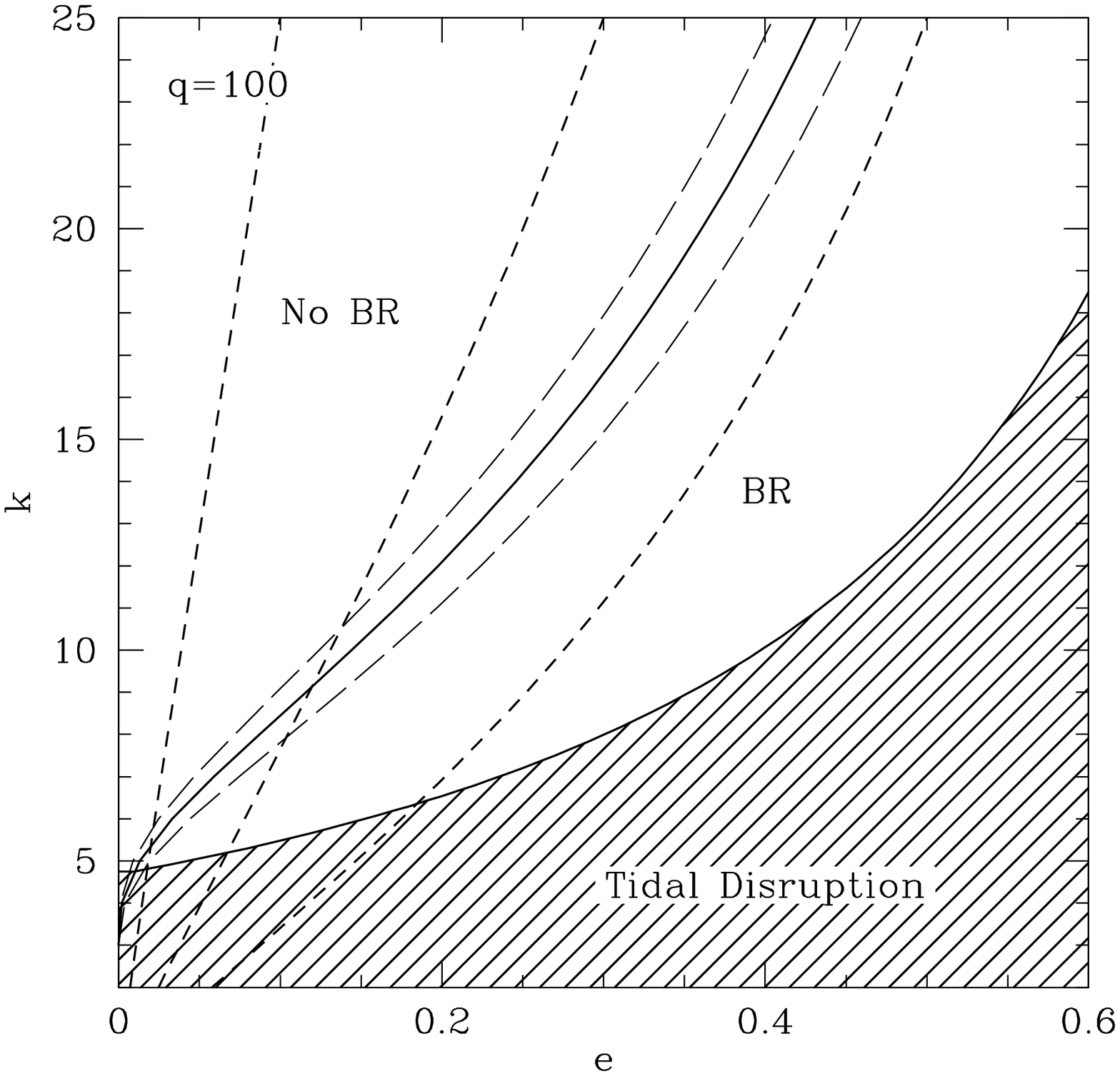} &
\includegraphics[width=2.65in]{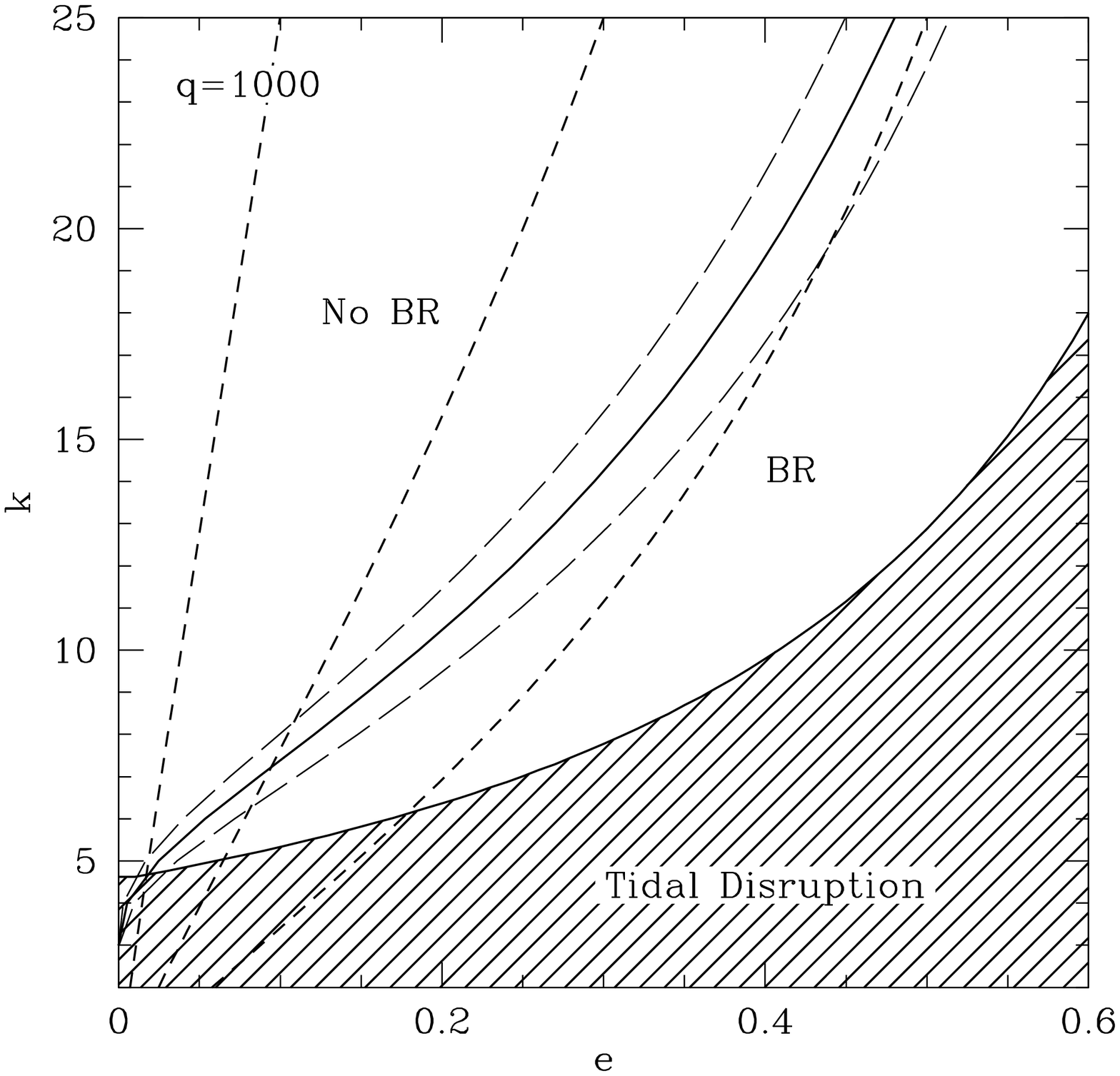} \\
\includegraphics[width=2.65in]{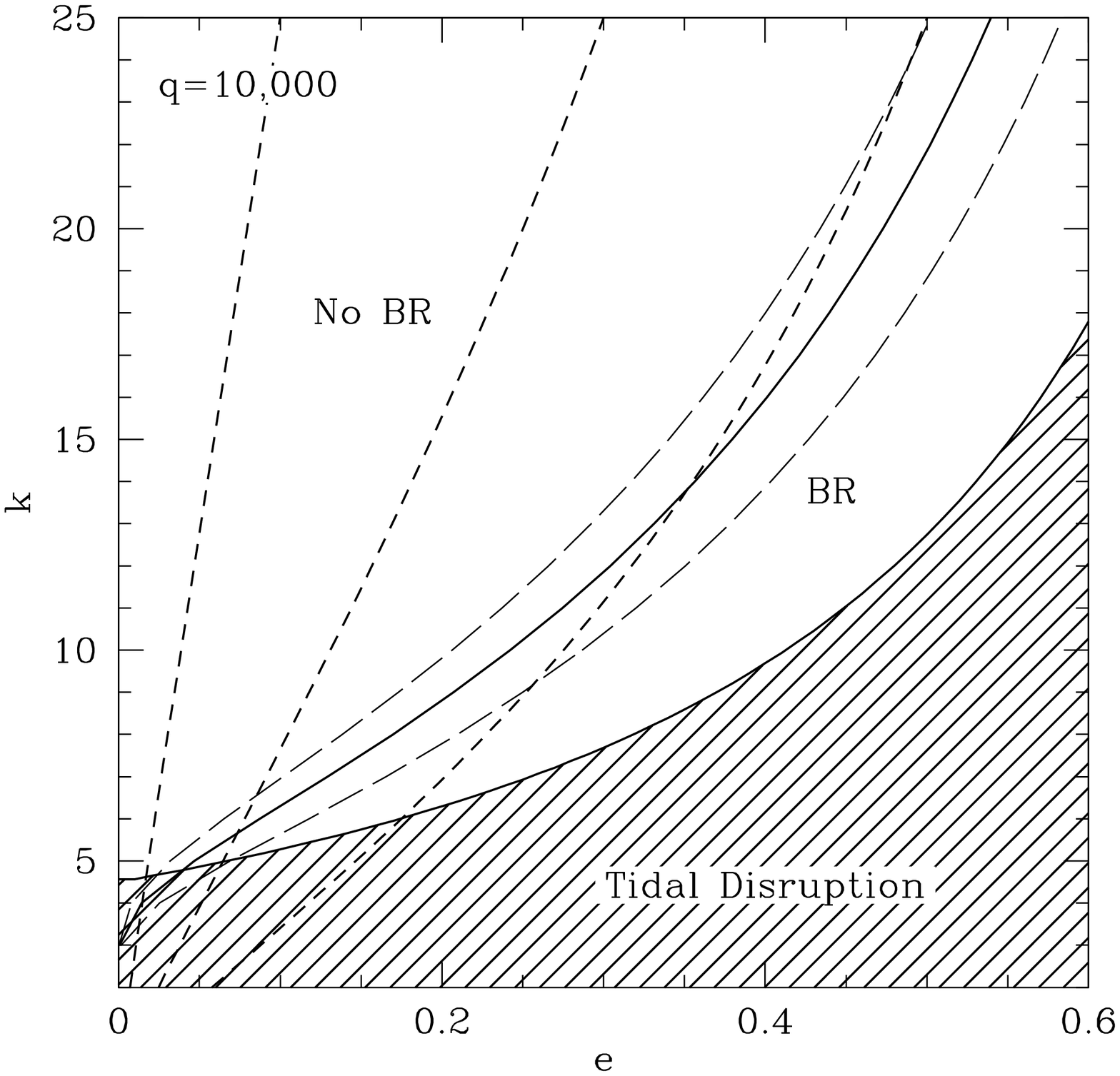} &
\includegraphics[width=2.65in]{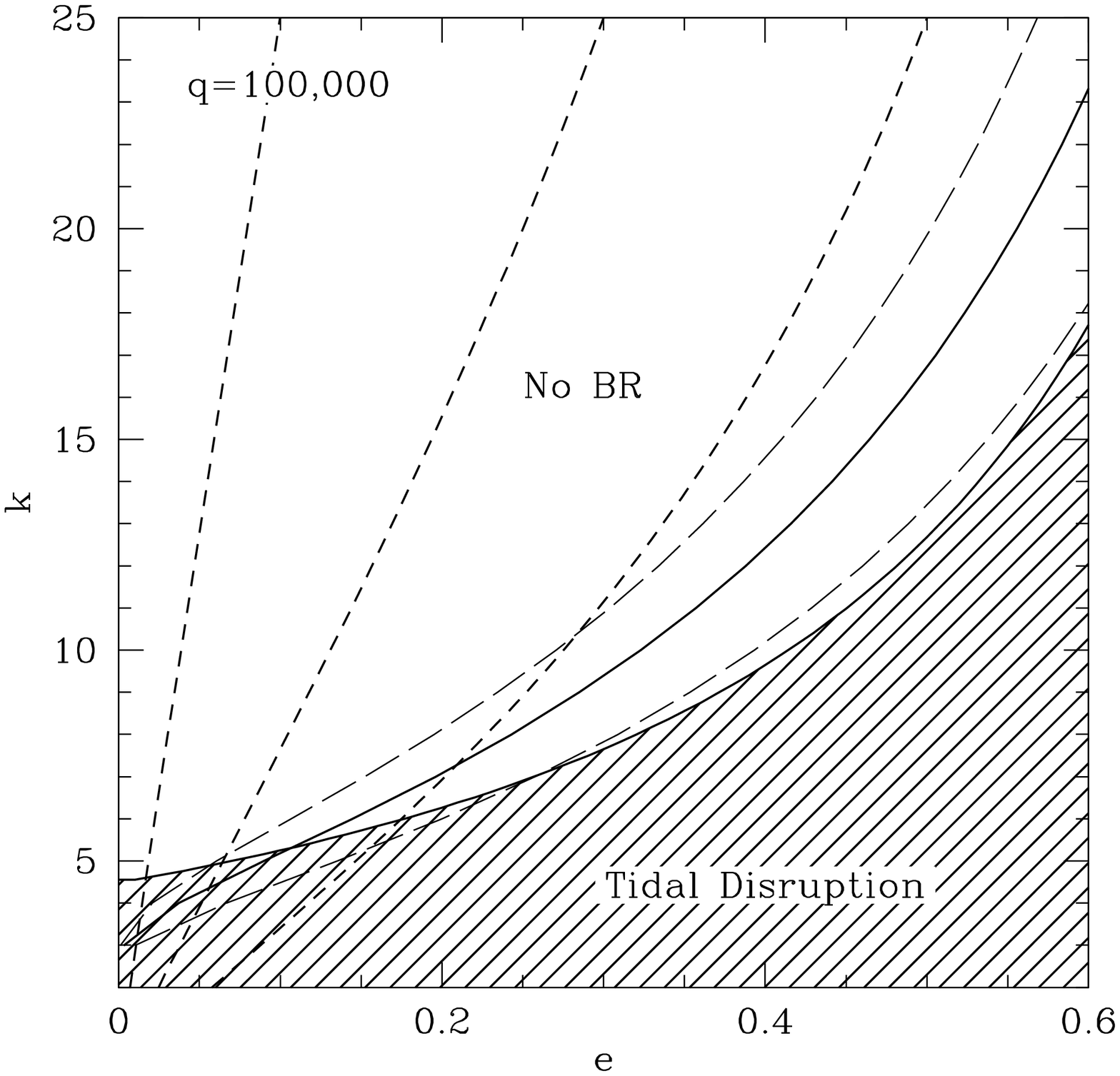}
\end{tabular}
\caption{\label{fig:limits} The regions in eccentricity-harmonic space where
back reaction is and is not important (labelled as ``BR'' and ``No BR'', respectively)
are delineated according to the $\chi_{jk}$ criterion for a $\ell =m=2$
\emph{f}-mode of a $0.6$~$M_\odot$ white dwarf,
and various mass ratios. In each plot, the solid curve traces out the contour
$\chi_{jk}=1$, and the long dashed lines to its left
and right trace $\chi_{jk}=0.1$ and $\chi_{jk}=10$, respectively.
The short dashed lines trace three gravitational radiation inspiral trajectories
through the plane. For reference, the tidal limit and the region where mode damping
via gravitational radiation during inspiral is important are also shown. Note that,
for the mode damping curves, the full, eccentricity-dependent inspiral times have
been used, rather than the circular orbit times.}
\end{figure*}

The delineation of the back reaction and no back reaction
regimes in the eccentricity-harmonic plane obtained with the above criterion for a
quadrupolar \emph{f}-mode on a $0.6$~$M_\odot$ white dwarf and various companion masses is
shown in Figure~\ref{fig:limits}. It is seen that the region of parameter space
where back reaction may be neglected, according to the $\chi_{jk}$ criterion,
grows with the companion mass. There is, however, a reason to think that back reaction
might actually play an important role in some regions of the parameter space where the $\chi_{jk}$
criterion indicates otherwise.

Consider the following thought experiment. Imagine that we are approaching
a resonance with an initial phase that would lead to a net negative energy
transfer in the no back reaction approximation. As we start removing
energy from the mode and depositing it into the orbit, the orbital frequency
will necessarily decrease (i.e.~the semi-major axis will increase) and
the system will get pushed off resonance. It will then have another chance to
approach the same resonance. Then, if the phase is such that energy is
transferred to the mode, then the system will once again get pushed off
resonance, but this time in the opposite direction (since the orbital frequency
will increase). Gravitational radiation will then evolve the system away from this
resonance and towards the next one. This scenario hints at the possibility that
back reaction may force the resonant energy transfer to be always positive.
However, this is not necessarily the case. For instance, we have assumed that
there is sufficient initial energy in the mode to be able to change the orbital
frequency significantly. Also, we have neglected the fact that gravitational
radiation will be removing energy from the orbit as we are transferring energy to
the orbit from the mode. If the rate of dissipation by gravitational radiation
is high enough, then back reaction may not matter.
The system will evolve through resonance regardless, on a time-scale determined by
the rate of dissipation. Hence, we can still get a net negative energy
transfer to the mode.

In summary, back reaction may be important in determining both
the magnitude and the direction of resonant energy transfer.
The $\chi_{jk}$ criterion provides, in some sense, only
a measure of the correction to the magnitude. In the regime
where $\chi_{jk}\gtrsim 1$, the implication is unambiguous:
back reaction will be essential in determining the energy
transfer. However, when $\chi_{jk}<1$, things are somewhat
uncertain for reasons stated above. A solution of the problem
including back reaction is required to determine conclusively
whether back reaction is important in that regime.

\subsection{Long term evolution} \label{SSEC:LTE}
In \S\ref{SEC:RET}, we calculated the energy transfer
for an individual resonance in the absence of back reaction. In general,
as the binary shrinks under gravitational radiation, the system will pass
through a sequence of resonances for each mode. However, this
is only a possibility for an eccentric orbit because, as demonstrated
previously, only the fundamental resonance exists for a circular orbit.
We note that, in the no back reaction approximation, the energy
transfer at a resonance can be negative as well as positive,
depending on the relative phase of the mode and the driver, and the initial amplitude.
Also, there will be negligible average energy transfer between resonances, as long
as we are well outside the tidal limit.
If we assume (as seems reasonable) that the system has no long term
phase memory, then the relative phasing at each resonance will be essentially
random, with a uniform distribution. It then follows that, on average, the mode
will tend to gain energy over time, and that the average total energy transfer
after a sequence of resonances will be simply the sum of the individual average
energy transfers given by (\ref{eq:nbr_ecc_en_tr}).

Let $\varepsilon_k$ denote the average energy transfer given by
(\ref{eq:nbr_ecc_en_tr}) for a particular mode at the $k$-th resonance,
and let $E_k$ be the energy (in units of $E_*$) in the mode before
the $k$-th resonance. It then follows from (\ref{eq:sho_delta_E})
and our assumptions of random phases and negligible energy transfer
between resonances that, for a sequence of resonances in the no
back reaction approximation, the evolution of the mode energy will
be given by the discrete random walk (with a drift)
\begin{equation} \label{eq:random_walk}
E_{k-1} = E_k + \varepsilon_k \left( 1 + 2\sqrt{\frac{E_k}{\varepsilon_k}}C_k \right) \ ,
\end{equation}
where $C_k$ is a random variable drawn from the distribution
\begin{equation*}
p(x) = \frac{1}{\pi\sqrt{1-x^2}} \ , \quad x\in [-1,1] \ .
\end{equation*}
For a derivation of elementary statistical properties of this random walk,
see Appendix~\ref{APP:RW}.

Figure~\ref{fig:mode_amp_nobr} shows the results from calculations of passage
through a sequence of resonances performed using the above random walk model
for several sets of initial conditions. We have chosen to plot the mode amplitude
\begin{equation}
B_j \equiv \frac{1}{R_{\ast}}\sqrt{\frac{2 E_j}{M_j \omega_j^2}} \ ,
\end{equation}
rather than the energy, because we want to draw attention to the fact that,
for moderate initial eccentricities, the amplitude of a $\ell=m=2$ \emph{f}-mode
can be driven to values in the range $\sim$0.1--1. (An amplitude of unity for
a $\ell=m=2$ mode corresponds to a maximum physical displacement of the stellar
surface of about 55\% relative to the unperturbed radius.)
We therefore expect that the linear normal mode analysis might not be valid in
those cases, and that non-linear effects may in fact determine the actual outcome.

\begin{figure*}
\begin{tabular}{cc}
\includegraphics[width=2.65in]{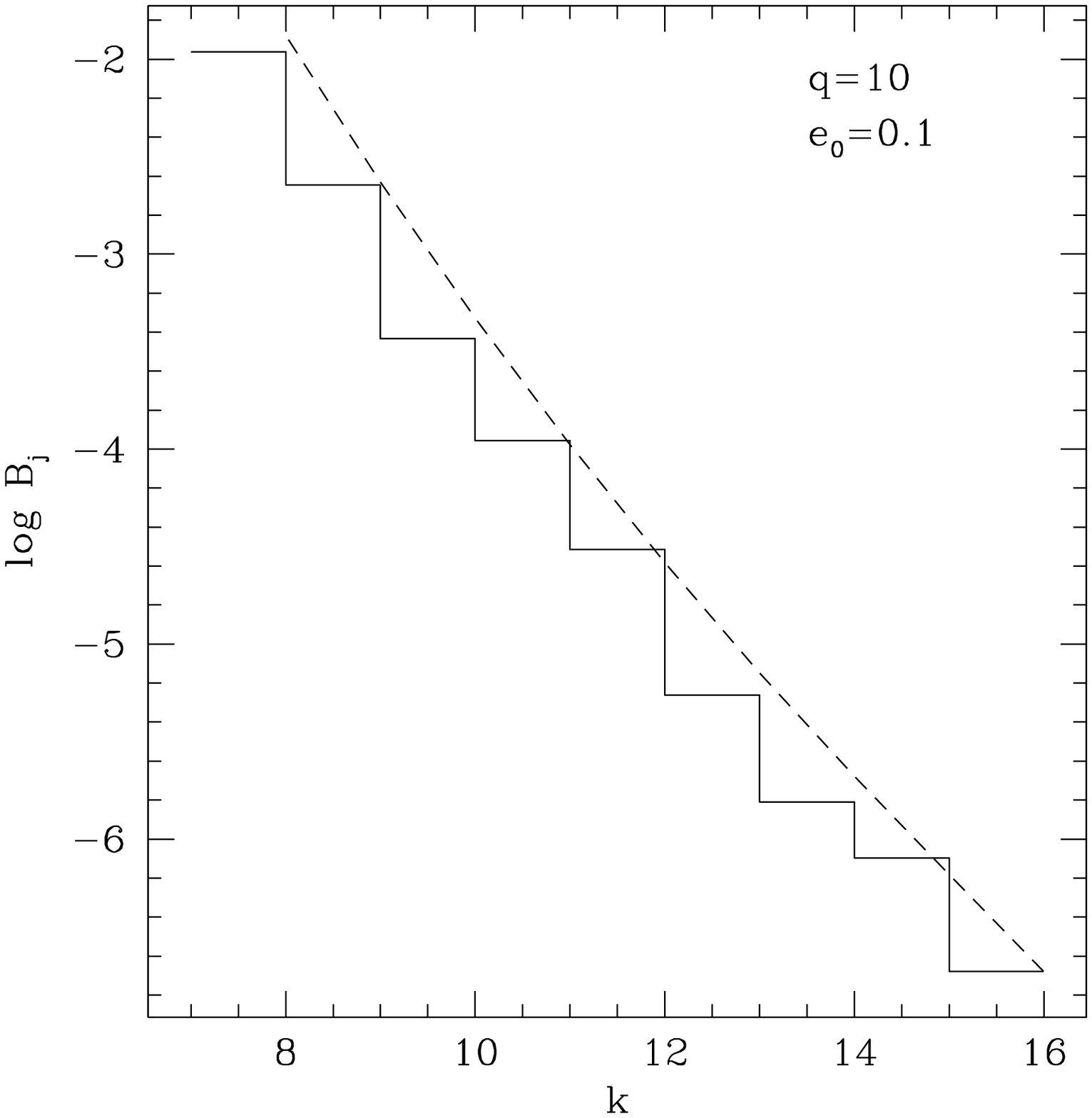} &
\includegraphics[width=2.65in]{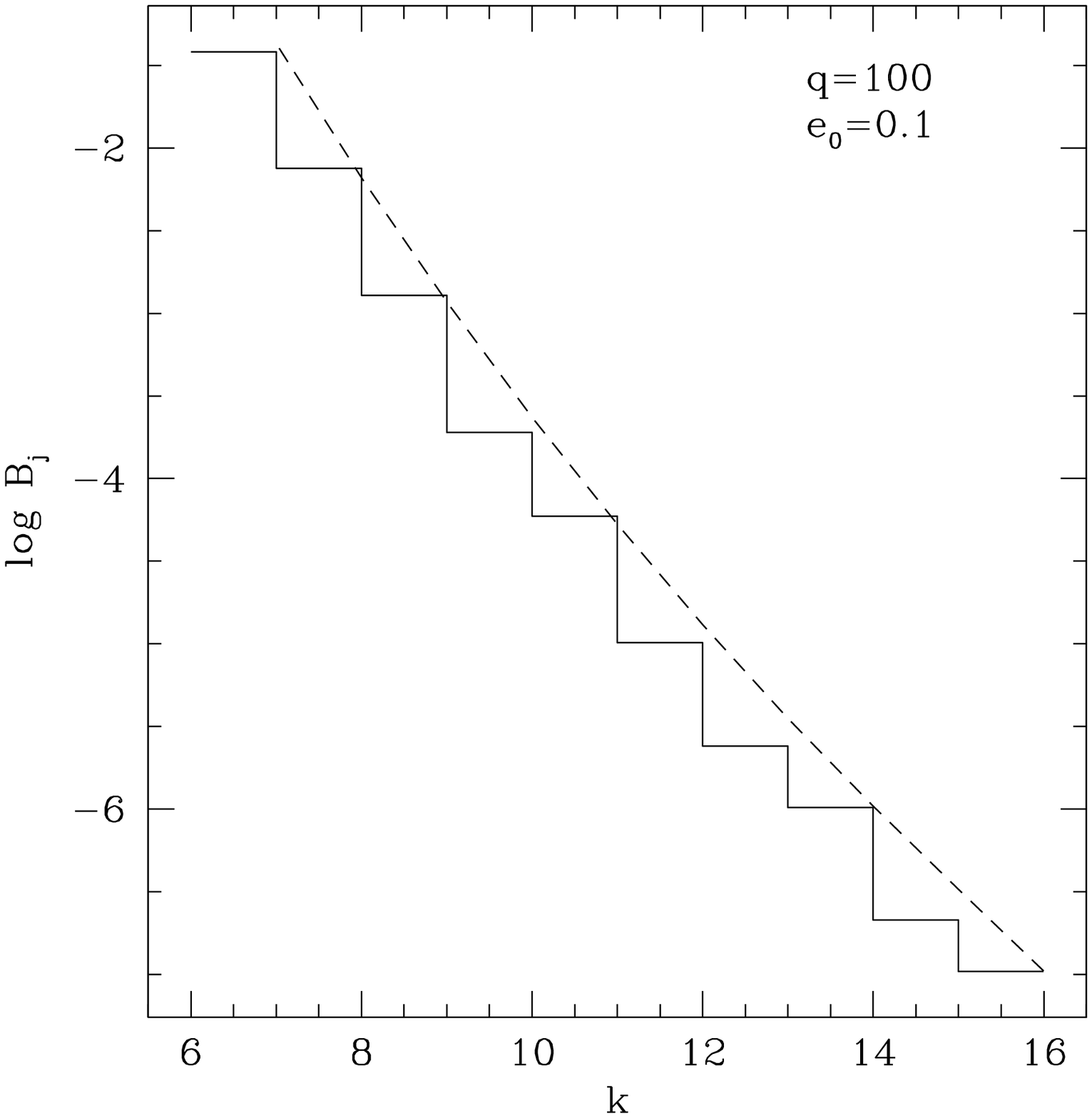} \\
\includegraphics[width=2.65in]{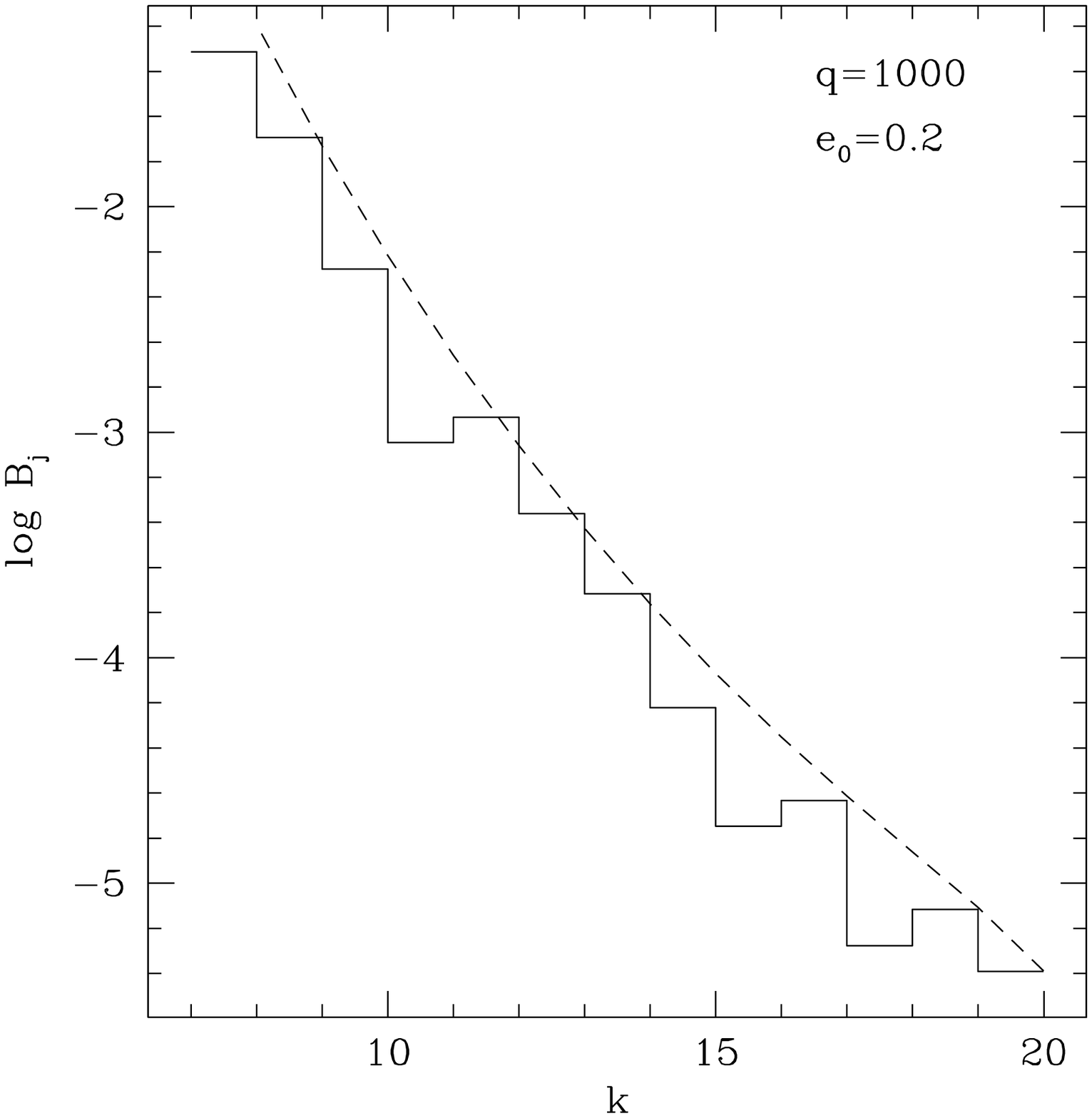} &
\includegraphics[width=2.65in]{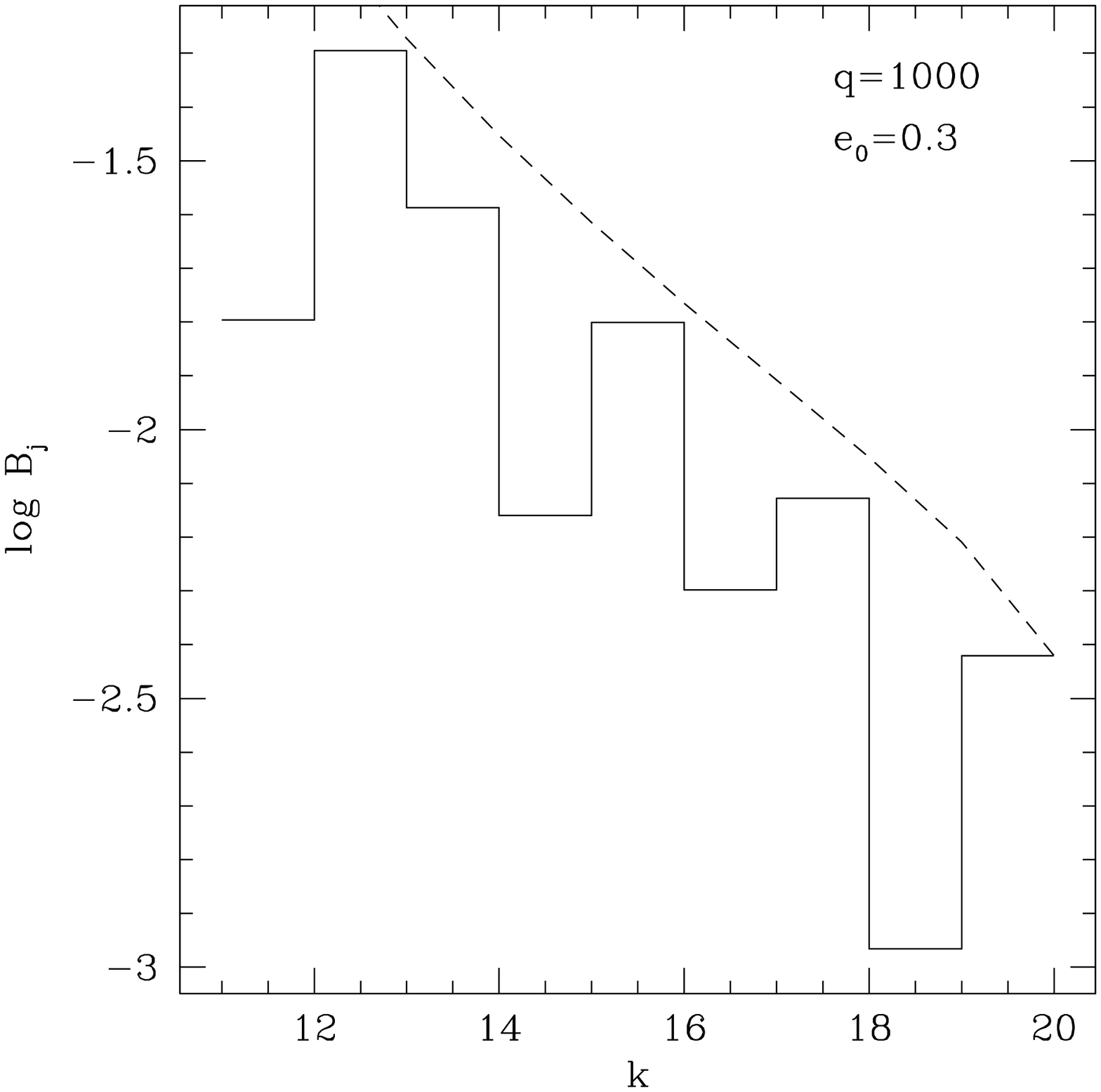} \\
\includegraphics[width=2.65in]{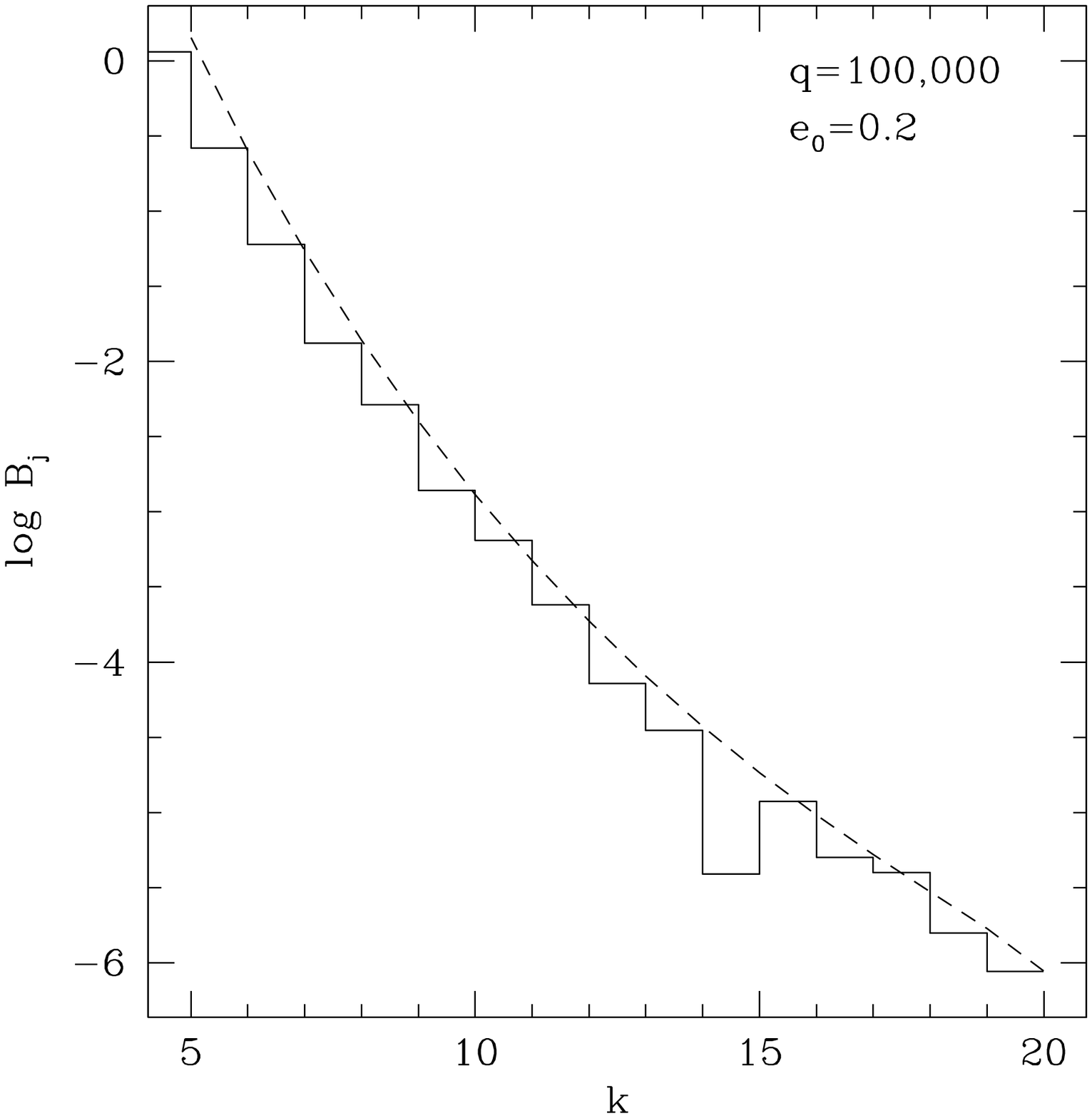} &
\includegraphics[width=2.65in]{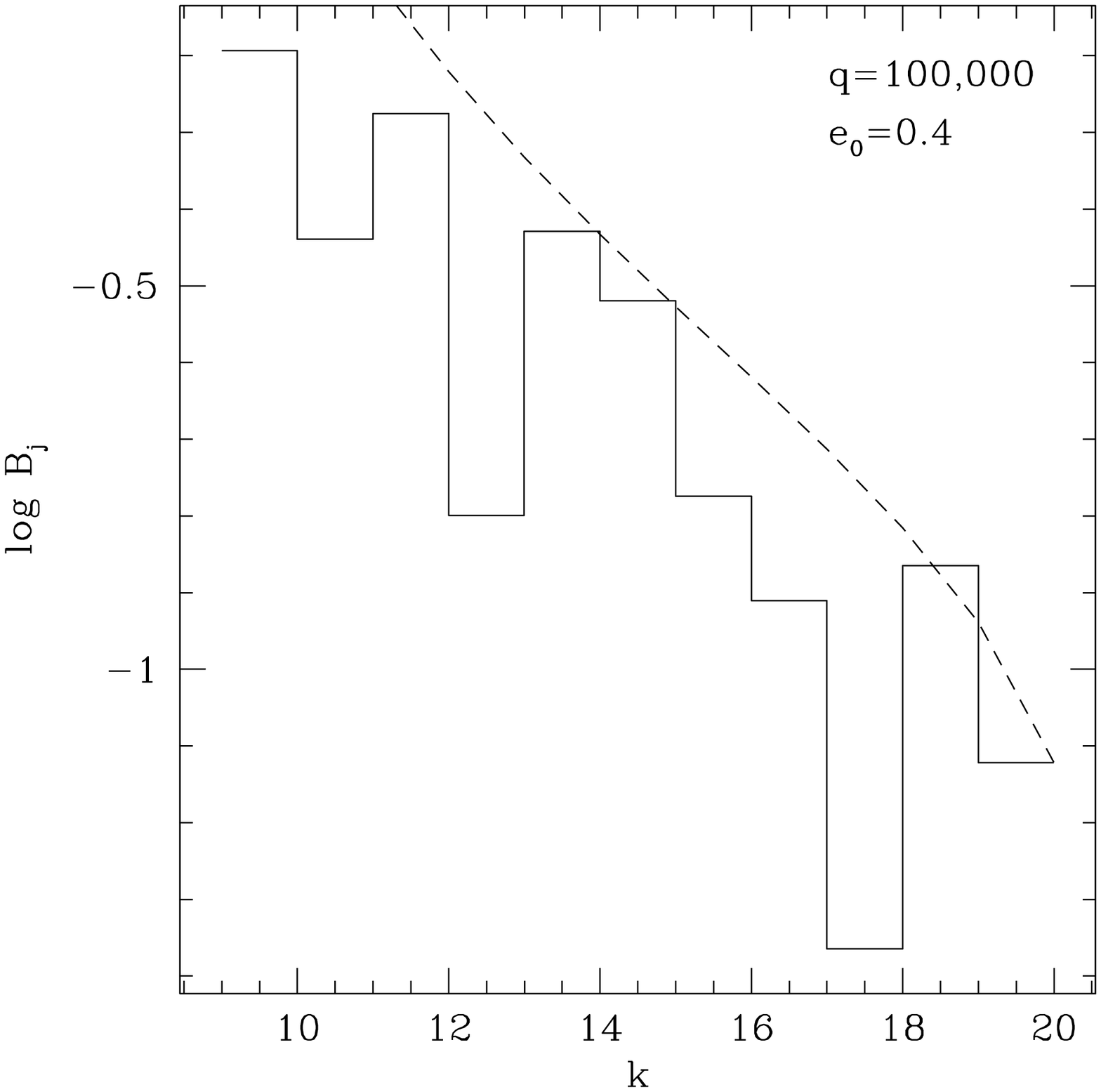}
\end{tabular}
\caption{\label{fig:mode_amp_nobr} The amplitude of a $\ell=m=2$ \emph{f}-mode of
a $0.6$~$M_\odot$ white dwarf during passage through a sequence of resonances in the
no back reaction approximation is shown for several sets of initial conditions.
All of these lie in regions of the eccentricity-harmonic plane where back reaction
is not important according to the $\chi_{jk}$ criterion. The calculations were done
using our semi-analytical formalism. In each case, the solid line shows a particular
realization of the random walk given by (\ref{eq:random_walk}), and the dashed line
follows the ensemble average. The random walks were terminated when $\chi_{jk}\sim0.01$.
Note that the scales on the axes are different for each plot.}
\end{figure*}

It should be noted that, even if back reaction plays a role in determining
the direction of energy transfer, our result that non-linear
amplitudes for a $\ell=m=2$ \emph{f}-mode can be attained
by passage through a sequence of resonances is unlikely to be affected.
This is due to the fact that the result depends chiefly upon the
allowed magnitude of energy transfer, and as we restricted our
calculations to the regime where $\chi_{jk}\ll 1$, back reaction
is not expected to change things.

\section{Conclusions}
In this paper, we have discussed a specific dynamical problem:
what is the energy transfer to the normal modes of a white
dwarf due to tidal resonances in a compact object binary
inspiralling under gravitational radiation?
For simplicity, we have considered only non-rotating,
completely degenerate stellar models.
We have provided a semi-analytical answer for the
case when the perturbation of the orbit by the modes may
be neglected (the no back reaction approximation). It has
been shown that, ignoring back reaction, and assuming that the
relative phase of the mode and the orbit is randomised between
successive resonances, the energy in a mode executes a discrete
random walk, with a drift, during passage through a sequence
of resonances.
In addition, we have found that the amplitude of a $\ell=m=2$
\emph{f}-mode can be driven into the non-linear regime as a result of
passage through such a sequence.
We have also demonstrated that back reaction can modulate
the amount of energy transferred significantly, depending upon the parameters
of the system. Our attempt to delineate the regime in parameter space
where the no back reaction approximation is valid has had mixed success:
we can predict when back reaction will be definitely important, but
a clear delineation of the regime where it will be unimportant has not been
found. Despite this ambiguity, our result that non-linear amplitudes
can be attained for a $\ell=m=2$ \emph{f}-mode is expected to be
robust. A treatment of the problem including back reaction will be the
subject of a future publication.

\section*{Acknowledgments}
RDB would like to thank J.~P.~Ostriker, K.~S.~Thorne, P.~Goldreich, Y.~Wu,
and M.~L.~Aizenmann for useful discussions and encouragement.
The authors acknowledge support under NASA grant 5-12032.

\bibliography{tidal_resonances1}

\appendix
\section{Details of Calculation for Forced Harmonic Oscillator} \label{APP:SHO}
From (\ref{eq:sho_zeta1}) and (\ref{eq:sho_acc}), it follows that
\begin{equation*}
\begin{split}
\zeta_1 = \frac{F_R}{2} e^{i\omega_0 \tau} \int_{\tau_0}^{\tau} &d\tau'\, \bigg\{\exp\left[i\left(\phi_R + \alpha\frac{\omega_0^2\tau'^2}{2}\right)\right] \\ &+ \exp\left[-i\left(\phi_R + 2\omega_0\tau' + \alpha\frac{\omega_0^2\tau'^2}{2}\right)\right]\bigg\} \ .
\end{split}
\end{equation*}
Completing the square in the argument of the second exponential, we get
\begin{equation*}
\begin{split}
\zeta_1 = \frac{F_R}{2} e^{i\omega_0 \tau} \Bigg\{ &I_1(\tau) \exp\left(i\phi_R\right) \\ &+ I_2(\tau) \exp\left[-i\left(\phi_R - \frac{2}{\alpha}\right)\right] \Bigg\} \ ,
\end{split}
\end{equation*}
where
\begin{align*}
I_1(\tau) &\equiv \int_{\tau_0}^{\tau} d\tau'\, \exp\left(\frac{i}{2}\alpha\omega_0^2\tau'^2\right) \ , \\
I_2(\tau) &\equiv \int_{\tau_0}^{\tau} d\tau'\, \exp\left[-\frac{i}{2}\left(\sqrt{\alpha}\omega_0\tau' + \frac{2}{\sqrt{\alpha}}\right)^2\right] \ .
\end{align*}
Making the change of variables
\begin{align*}
u(\tau') &\equiv \sqrt{\frac{\alpha}{\pi}}\omega_0\tau' \ , \\
v(\tau') &\equiv \sqrt{\frac{\alpha}{\pi}}\omega_0\tau' + \frac{2}{\sqrt{\pi\alpha}} \ ,
\end{align*}
we have
\begin{align*}
I_1 &= \frac{1}{\omega_0}\sqrt{\frac{\pi}{\alpha}}\int_{u_0}^{u_\tau} du\, \exp\left(\frac{i\pi u^2}{2}\right) \ , \\
I_2 &= \frac{1}{\omega_0}\sqrt{\frac{\pi}{\alpha}}\int_{v_0}^{v_\tau} dv\, \exp\left(-\frac{i\pi v^2}{2}\right) \ ,
\end{align*}
where $u_0 = u(\tau_0)$, $u_\tau = u(\tau)$, et cetera.
Evaluating the integrals gives
\begin{align*}
I_1 &= \frac{1}{\omega_0}\sqrt{\frac{\pi}{\alpha}} \Big\{ C(u_\tau)-C(u_0) + i\left[S(u_\tau)-S(u_0)\right]\Big\} \ , \\
I_2 &= \frac{1}{\omega_0}\sqrt{\frac{\pi}{\alpha}} \Big\{ C(v_\tau)-C(v_0) - i\left[S(v_\tau)-S(v_0)\right]\Big\} \ ,
\end{align*}
where the Fresnel integrals are defined by
\begin{align*}
C(z) &\equiv \int_0^z dt\, \cos\left(\frac{\pi t^2}{2}\right) \ , \\
S(z) &\equiv \int_0^z dt\, \sin\left(\frac{\pi t^2}{2}\right) \ .
\end{align*}
Note that
\begin{equation*}
\lim_{z\rightarrow\pm\infty} C(z) = \lim_{z\rightarrow\pm\infty} S(z) = \pm\frac{1}{2}
\end{equation*}
\citep[see, for example,][]{abr72}.
Let us now consider the energy after long times.
If we start out from zero frequency, then $\omega_0\tau_0= -\alpha$, and
\begin{align*}
u_0 &= -\frac{1}{\sqrt{\pi\alpha}} \ll -1 \ , & v_0 &= \frac{1}{\sqrt{\pi\alpha}} \gg 1 \ .
\end{align*}
The $I_2$ integral will not contribute significantly since its upper and lower
limits are effectively the same.
For $I_1$, we find
\begin{equation*}
I_1 = \frac{1}{\omega_0}\sqrt{\frac{\pi}{\alpha}}\left(1+i\right) = \frac{1}{\omega_0}\sqrt{\frac{2\pi}{\alpha}} e^{i\pi/4} \ .
\end{equation*}
Hence, after some algebra,
\begin{equation*}
\Delta E \equiv E - E_0 = \frac{\pi F_R^2}{4\alpha\omega_0^2}\left( 1 + 2\sqrt{\frac{E_0}{\pi F_R^2/4\alpha\omega_0^2}}\cos\psi\right) \ ,
\end{equation*}
where $E_0=\left|\zeta_0\right|^2/2$ and $\psi$ is an initial phase.

\section{Hansen Coefficients} \label{APP:HC}
The Hansen coefficients $X_k^{p,m}$ for the two-body problem are defined by
\begin{equation*}
\left(\frac{R}{a}\right)^p \exp(imv) = \sum_{k= -\infty}^{\infty} X_k^{p,m}(e) \exp(ikl) \ ,
\end{equation*}
where $R$ is the orbital separation, $a$ is the semi-major axis, $v$ is the
true anomaly, $e$ is the orbital eccentricity, and $l$ is the mean anomaly.
The Hansen coefficients are real functions of the eccentricity, and
it can be shown that, to lowest order in eccentricity,
\begin{equation*}
X_k^{p,m}(e) \propto e^{\left|k-m\right|}
\end{equation*}
\citep[][and references therein]{mur99}. The coefficients can be calculated
to any desired order in eccentricity as a series in terms of Newcomb operators:
\begin{equation*}
X_k^{p,m}(e) = e^{\left|k-m\right|}\sum_{\nu=0}^{\infty}X^{p,m}_{\nu+\lambda,\nu+\zeta} e^{2\nu} \ ,
\end{equation*}
where $\lambda=\max(0,k-m)$, $\zeta=\max(0,m-k)$, and the Newcomb operators
$X^{a,b}_{c,d}$ are defined via recursion relations
\citep[see][]{mur99}. Alternatively, for quantitative work, the Hansen coefficients
can be evaluated for a given eccentricity by calculating the integral
\begin{equation*}
X_k^{p,m}(e) = \frac{1}{2\pi}\int_0^{2\pi} dl\, \left(\frac{R}{a}\right)^p \cos(mv-kl)
\end{equation*}
numerically.

\section{Damping of Quadrupolar Modes Under Gravitational Radiation} \label{APP:DOMUGR}
The average power radiated in gravitational waves due to a time-dependent mass quadrupole
moment is given, in the weak-field limit of general relativity, by
\begin{equation*}
\frac{d E_{\rm GW}}{d t} = \frac{G}{45 c^5} \langle\dddot{Q}_{i j}\dddot{Q}_{i j}\rangle
\end{equation*}
\citep{mis73}, where $Q_{i j}$ is the mass quadrupole moment as defined conventionally
in classical physics:
\begin{equation*}
Q_{i j} = \int d^3x\, \left( 3 x_i x_j - r^2 \delta_{i j}\right) \rho(\mathbf{x},t) \ .
\end{equation*}
It can be shown that the power radiated by a quadrupolar mode is independent of $m$
(this is a consequence of the Wigner-Eckart theorem). Hence, we can restrict ourselves
to the $m=0$ case for simplicity.
It then follows that $Q_{11}=Q_{22}= -Q_{33}/2$, and that the
off-diagonal terms vanish. Therefore, noting that the time dependence is sinusoidal,
we have
\begin{equation}\label{eq:gw_mode}
\frac{d E^{\rm GW}_j}{d t} = -\frac{G}{60 c^5} \omega_j^6 \dddot{Q}_{33}^2 \ ,
\end{equation}
where $Q_{33}$ should now be understood to mean the time-independent amplitude of the
mass quadrupole moment. Writing the mass density as
\begin{equation*}
\rho(\mathbf{x}) = \rho_0(r) + B_j \delta\rho_j(r) Y_{20}(\mathbf{\hat{r}}) \ ,
\end{equation*}
where $B_j$ is the amplitude of the mode and $\delta\rho_j$ is the
normalised density perturbation associated with the mode, we find
\begin{equation*}
Q_{33} = 4\sqrt{\frac{\pi}{5}} B_j \int_{0}^{R_\ast} dr\, r^4 \delta\rho(r) \ .
\end{equation*}
The above integral can be simplified by using the linearised Poisson equation,
integrating by parts twice, and using the surface boundary condition
$\eta_4 = -(\ell+1)\eta_3$. The result is
\begin{equation*}
Q_{33} = \sqrt{\frac{5}{\pi}} M_{\ast} R_{\ast}^2 \eta_{3j}(R_{\ast}) B_j \ .
\end{equation*}
Substituting the above expression into (\ref{eq:gw_mode}), we get
\begin{equation*}
\frac{d E^{\rm GW}_j}{d t} = -\frac{G M_{\ast}^2 R_{\ast}^4}{12\pi c^5} \eta_{3j}^2(R_{\ast}) \omega_j^6 B_j^2 \ .
\end{equation*}
Finally, noting that the total energy for an isolated mode is given by
\begin{equation*}
E_j = \frac{1}{2} M_j R_{\ast}^2 \omega_j^2 B_j^2 \ ,
\end{equation*}
we arrive at the $e$-folding time for the mode energy under damping by gravitational radiation:
\begin{equation*}
T_j = \frac{6\pi}{\omega_{\ast}}\beta_{\ast}^{-5} \eta_{3j}^{-2}(R_{\ast})\left(\frac{M_j}{M_{\ast}}\right) \sigma_j^{-4} \ .
\end{equation*}

\section{Statistical Properties of Resonant Energy Transfer in the No Back Reaction Approximation} \label{APP:RW}
In this appendix, we derive some statistical properties of the random walk
given by (\ref{eq:random_walk}), which we rewrite as
\begin{equation} \label{eq:random_walk2}
E_{k-1} = E_k + \varepsilon_{k} + Z_{k} \ ,
\end{equation}
where
\begin{equation*}
Z_k \equiv 2\sqrt{\varepsilon_k E_k} C_k \ .
\end{equation*}
Recall that $\varepsilon_k$ is known in advance, and that $C_k$ is a random
variable drawn from the distribution
\begin{equation*}
p(x) = \frac{1}{\pi\sqrt{1-x^2}} \ , \quad x \in [-1,1] \ .
\end{equation*}
Note that
\begin{equation*}
\langle C_k \rangle = 0 \ , \quad \langle C_k^2 \rangle = \frac{1}{2} \ .
\end{equation*}
Since $E_k$ will only depend upon $C_\alpha$, for $\alpha>k$, $Z_k$ is linear
in $C_k$. As all the $C_k$ are independent random variables (by assumption),
it follows that
\begin{equation} \label{eq:z_moments}
\langle Z_k \rangle = 0 \ , \quad \langle Z_k^2 \rangle = 2\varepsilon_k\langle E_k \rangle \ .
\end{equation}

Given an initial mode energy $E_k$ before the $k$-th resonance, we wish to determine
the average mode energy $\langle E_{k-p} \rangle$, and its variance $\sigma_p^2$,
after passage through $p$ resonances.
From (\ref{eq:random_walk2}), we have
\begin{equation} \label{eq:random_walk3}
E_{k-p} = E_{k} + \sum_{\alpha=k-p+1}^{k}\varepsilon_{\alpha} + \Delta_p \ ,
\end{equation}
where we have defined
\begin{equation*}
\Delta_p \equiv \sum_{\alpha=k-p+1}^{k} Z_{\alpha} \ .
\end{equation*}
Using (\ref{eq:z_moments}), it follows immediately that
\begin{equation} \label{eq:rw_avg_en}
\langle E_{k-p} \rangle = E_k + \sum_{\beta=k-p+1}^k \varepsilon_{\beta} \ .
\end{equation}
Note that, since $\varepsilon_{\beta}>0$, $\langle E_{k-p} \rangle$ increases
monotonically as we pass through a sequence of resonances. Hence, we say that
the random walk (\ref{eq:random_walk2}) has a drift.
To calculate the variance, we need to find $\langle E_{k-p}^2 \rangle$. Writing
\begin{equation*}
\begin{split}
E_{k-p}^2 = \left(E_k + \sum_{\alpha=k-p+1}^k \varepsilon_{\alpha}\right)^2 + \Delta_p^2 + O(\Delta_p) \ ,
\end{split}
\end{equation*}
we see that
\begin{equation} \label{eq:rw_en_var1}
\langle E_{k-p}^2 \rangle = \langle E_{k-p} \rangle^2 + \langle \Delta_p^2 \rangle \ ,
\end{equation}
as all the terms linear in $\Delta_p$ will vanish when averaged. To calculate
$\langle \Delta_p^2 \rangle$, we write
\begin{equation*}
\Delta_p^2 = \sum_{\alpha=k-p+1}^k Z_{\alpha}^2 + 2\sum_{\alpha=k-p+1}^{k-1}\sum_{\beta=\alpha+1}^k Z_{\alpha}Z_{\beta} \ ,
\end{equation*}
and note that, since $\beta>\alpha$ in the above double sum, each term of the double sum
will be linear in $C_{\alpha}$, and will, hence, vanish upon averaging. Therefore, we have
\begin{equation*}
\langle\Delta_p^2\rangle = 2\sum_{\alpha=k-p+1}^k \varepsilon_{\alpha}\langle E_{\alpha} \rangle \ .
\end{equation*}
Substituting into (\ref{eq:rw_en_var1}), and then using (\ref{eq:rw_avg_en}), we find
\begin{equation}
\sigma_p^2 = 2\sum_{\alpha=k-p+1}^k \varepsilon_{\alpha}\left( E_k + \sum_{\beta=\alpha+1}^k \varepsilon_{\beta} \right) \ .
\end{equation}

It should be noted that (\ref{eq:random_walk2}) is not Gaussian, nor will it become Gaussian
after many resonances. That the process is not Gaussian is clear from the fact that the
random walk is bounded from below. Furthermore, the central limit theorem is not applicable
because, typically, the probability distribution of $\Delta_p$ is dominated by the most recent
few harmonics, and hence the effective number of variables never becomes large.

\bsp
\label{lastpage}
\end{document}